\documentclass[prd,preprint,tightenlines,floatfix,showpacs,preprintnumbers,nofootinbib,eqsecnum]{revtex4-1}

\pdfoutput=1

\usepackage[dvips,final]{graphicx}
\usepackage{color,xcolor}
\usepackage{amssymb}
\usepackage{amsmath}
\usepackage{amsfonts}
\usepackage{epsfig}
\usepackage{bm}
\usepackage{comment}
\usepackage{float}
\usepackage[utf8]{inputenc}
\usepackage{comment}

\colorlet{RED}{red}

\def\qq{$q\bar{q}$ }

\DeclareSymbolFont{myletters}{OML}{ztmcm}{m}{it}
\DeclareMathSymbol{\uplambda}{\mathord}{myletters}{"15}

\begin{document}
 
% --------------------------------------------------------------
%                         Start here
% --------------------------------------------------------------

\title{\textbf{Coherent photoproduction of light vector mesons off nuclear targets in the dipole picture}}

\author{Cheryl Henkels$^{1}$}
\email{cherylhenkels@hotmail.com}

\author{Emmanuel G. de Oliveira$^{1}$}
\email{emmanuel.de.oliveira@ufsc.br}

\author{Roman Pasechnik$^{2}$}
\email{Roman.Pasechnik@hep.lu.se}

\author{Haimon Trebien$^{1}$}
\email{haimontrebien@outlook.com}

\affiliation{
\\
{$^1$\sl Departamento de F\'isica, CFM, Universidade Federal 
de Santa Catarina, C.P. 5064, CEP 88035-972, Florian\'opolis, 
SC, Brazil
}\\
{$^2$\sl
Department of Physics, Lund
University, SE-223 62 Lund, Sweden\vspace{5mm}
}}

\begin{abstract}\vspace{5mm}
We study the coherent photoproduction of light vector mesons in Pb-Pb collisions in the framework of color dipole approach. We employ the Glauber--Gribov formalism supplemented by an effective suppression factor $R_G$ accounting for the gluon shadowing correction. We adjust the latter to reproduce the deep inelastic structure function $F_2$ (E665) and $\rho$ meson photoproduction (ALICE) data. We achieve a good description of the available data points with $R_G = 0.85$ at scale $M_\rho^2/4 = 0.15$ GeV$^2$. In addition, employing this suppression factor, we present predictions for coherent $\rho(2S)$, $\omega(1S,2S)$ and $\phi(1S,2S)$ photoproduction observables using the holographic vector meson wave functions.
\end{abstract}

\maketitle  

\newpage

%----------------------------------------------------------
\section{Introduction}

During the last few years, the LHC collaborations have been publishing new experimental data for vector meson photoproduction in PbPb ultraperipheral collisions (UPCs). Complementing RHIC data~\cite{STAR:2002caw, STAR:2011wtm, PHENIX:2009xtn}, LHC provides measurements of both heavy and light states, such as $J/\Psi$ \cite{CMS:2016itn, ALICE:2012yye, ALICE:2013wjo, ALICE:2015nmy, ALICE:2023svb} and $\rho$ \cite{ALICE:2015nbw, ALICE:2020ugp}. In the latter case, non-perturbative effects are enhanced due to a small $\rho$-meson mass making a reliable description of this process more difficult to achieve. The lack of knowledge about the mechanisms that govern coherent photoproduction of light vector mesons is one of the main motivations for this work. 

In exclusive photoproduction processes at high-energy nucleus-nucleus UPCs the two nuclei scatter at impact parameters larger than the sum of their radii. This interaction can be described by the color dipole model, where a real photon (with virtuality $Q^2 = 0$), originating from the projectile nucleus, fluctuates into a $q\bar q$ dipole, which then interacts with the target nucleus via the exchange of low-$x$ gluons. Since in such a scattering, the quantum numbers of the target nucleus remain unchanged, one often refers to it as due to an exchange of a colorless object called Pomeron \cite{Jenkovszky:2022qnc,Nikolaev:1994kk, Mueller:1994jq}.

As a continuation of our previous work on light vector mesons \cite{Henkels:2022bne} produced in $\gamma p$ interactions to nuclear targets, here we aim at computing the differential (in rapidity) cross sections of the three light vector mesons, $\rho$, $\omega$, and $\phi$ using the color dipole model~\cite{Nikolaev:1990ja,Nikolaev:1991et}. For the proton target case, a satisfactory description of data was obtained, despite the fact that the light vector meson production is a soft process, without a hard scale. However, it is well known that the color dipole model works even in the absence of a perturbative scale, such as electron--proton deep inelastic scattering at very small virtualities. One supporting argument is that, when multiple gluons are exchanged in the soft domain at high energies, the saturation scale plays the role of a hard scale and justifies the perturbative approach. This scale sets the boundary when gluons start to recombine; it is known to be larger in the nucleus than in the proton, further supporting the applicability of the color dipole model.

Note, the exclusive cross sections are in general very sensitive to the modeling of the color dipole interaction with the proton \cite{Henkels:2020qvo,Goncalves:2022wzq, Cepila:2019skb}. Such sensitivity comes mostly from soft and non-perturbative kinematic regimes~\cite{Kopeliovich:2001xj} poorly constrained by traditional fits of the dipole parameterizations to the hard Deep Inelastic Scattering (DIS) data from the HERA collider. In this work, we show the numerical results obtained with the latest DIS fit of Ref.~\cite{Golec-Biernat:2017lfv} for the Golec-Biernat--Wuesthoff (GBW) dipole model \cite{Golec-Biernat:1998zce,Golec-Biernat:1999qor}. The latter is the simplest well-established dipole parameterization known for its success in description of the wealth of experimental data on both inclusive and exclusive production processes, which includes both hard and soft scales processes, in lepton-proton ($lp$), $\gamma \gamma$ interaction, proton-proton ($pp$), proton-nucleus ($pA$) and nucleus-nucleus ($AA$) collisions, e.g., Refs. \cite{Timneanu:2001bk, Henkels:2022bne, Forshaw:2003ki, Flensburg:2008ag, Henkels:2020kju, Goda:2023jie, Albacete:2010sy, Goncalves:2017chx, Luszczak:2013rxa, Bartels:2002uf}.

We will be using the AdS/QCD holographic wave functions~\cite{Brodsky:2008pg,Brodsky:2014yha,Forshaw:2012im}, the same ones used in our previous paper \cite{Henkels:2022bne}, where the photoproduction cross section on the proton target was well described. The holographic approach enables us to calculate the differential photoproduction cross sections for vector mesons in both ground and excited states, which is a difference of the formalism in contrast to the usual boosted Gaussian \cite{Forshaw:2003ki} and the Gauss-LC \cite{Kowalski:2003hm,Frankfurt:1997fj} wave functions, where the parameters are fitted to vector meson production data. Indeed, such analyses involving the excited mesons are still scarce in the literature, and a full self-consistent description of exclusive observables is lacking. The advantage of the holographic approach is that the solutions for the relativistic light-front (LF) equation are frame independent \cite{RevModPhys.21.392}. Since the amplitudes in the dipole model are also calculated in the LF frame, the hadron wave functions obtained with the holographic approach are very convenient for practical analysis. 

In our previous works devoted to vector meson exclusive photoproduction off nuclear targets \cite{Henkels:2020kju,Henkels:2020qvo}, we observed that a suppression of the total dipole-nucleus cross section in comparison with the dipole-nucleon one should be accounted for in order to describe the available data for $J/\Psi$ and $\psi(2S)$ final states. Such a suppression is due to the so-called gluon shadowing phenomenon when the gluon density in a free nucleon appears to be different compared to that in a nucleon bound inside a nucleus. As was demonstrated earlier in Refs.~\cite{Henkels:2020kju,Henkels:2020qvo} the gluon shadowing accounted for through the use of a nuclear parton density function (PDF) for gluons at small $x$ was crucial for description of the available data on heavy vector mesons photoproduction, such as $J/\Psi$ and $\psi(2S)$. In this work, the gluon shadowing must be extracted from data, since nuclear PDFs are not known at soft factorization scales typical for light states, such as the $\rho$ meson. Besides the latter, below we have also studied the coherent photoproduction of $\rho(2S)$, $\omega(1S,2S)$ and $\phi(1S,2S)$ states. 

For the light vector meson states, motivated by the analysis of Ref.~\cite{Brodsky:2014yha}, we consider the $\rho (770)$ and $\rho (1450)$ as the ground and excited states of the $\rho$ meson, respectively. For $\omega$ mesons, we consider $\omega (782)$ and $\omega(1420)$ as the ground and excited states, respectively. And for $\phi$, we have $\phi(1020)$ and $\phi(1680)$ as ground and excited states, respectively. All of the above
states are listed on the PDG~\cite{ParticleDataGroup:2016lqr}.

Alternatively to the color dipole model, the vector meson dominance model (VMD) consists of modeling the real photon as a superposition of a bare photon state with vector meson states. This model has been very successful in describing the photoproduction of 1S states~\cite{Frankfurt:1997zk, Frankfurt:2015cwa, Khoze:2019xke}. As our intention here is to broaden the applicability of the color dipole model and describe excited states, we do not pursue this alternative.

The article is organized as follows. In Sect.~\ref{formalism} we provide a brief discussion of the nuclear vector-meson photoproduction process and relevant observables in terms of the holographic LF meson wave functions and the universal dipole cross section. Sect.~\ref{results} presents our numerical results for the gluon shadowing effect, extracted effectively from the available data on the $F_2^A$ structure function, and for $\rho$ photoproduction cross section. Considering the gluon shadowing as a global suppression factor for the light vector mesons photoproduction observables, we make predictions for the ground and excited states of $\rho$, $\omega$ and $\phi$ mesons. At last, a brief summary of our results is given in Sect.~\ref{conclusion}.

%--------------------------------------------
\section{Coherent vector meson photoproduction formalism}
\label{formalism}
%--------------------------------------------

\begin{figure}[!h]
\centering
\includegraphics[width = 14cm]{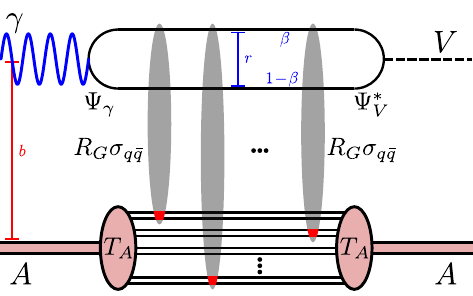}
\caption{A schematic illustration of the amplitude for the coherent vector ($V$) meson photoproduction process in $\gamma A\to V A$ scattering. The relevant kinematic variables here are the impact parameter $\bm{b}$ being the transverse separation of the photon from the center of the target, the \qq transverse separation $\bm{r}$, and the photon momentum fraction carried by the quark $\beta$. The effective vertices of the photon splitting into the \qq pair and the \qq pair recombining into the vector meson $V$ are represented, respectively, by $\Psi_\gamma$ and $\Psi^*_V$ LF wave functions. The gray bubbles represent the QCD interactions of the dipole with different nucleons inside the target nucleus, which are described by the universal dipole cross section $\sigma_{q\bar q}$ corrected for the gluon shadowing effect by the $R_G$ factor. At last, $T_A$ is the thickness function as discussed in the text.}
\label{fig:diagram}
\end{figure}

In the color dipole picture, the exclusive coherent photoproduction processes in $AA$ UPCs occur at large impact parameters through a quasi-real photon radiation off the projectile nucleus, with a subsequent photon fluctuation into a $q\bar q$ dipole that elastically scatters off the target nucleus. While the $q\bar q$ dipole effectively ``probes'' the inner structure of the target via strong interactions, no such interactions occur between the nuclei themselves, and both of them are left intact in the final state. Besides, for the process of our interest here, the $q\bar{q}$ dipole hadronizes into a given vector meson state as illustrated in Fig.~\ref{fig:diagram}.
%%%%%%%%%%%%%%%%%%%%%%%%%%%%%%%%%%%%%%%%%%%%%%%%%%%%%%%%%%%%%%%
%%%%%%%%%%%%%%%%%%%%%%%%%%%%%%%%%%%%%%%%%%%%%%%%%%%%%%%%%%%%%%%

The cross section for this process is given by~\cite{Ivanov:2002kc,Ivanov:2002eq}:
\begin{align} \nonumber 
\frac{d\sigma_{A A\rightarrow AV A}}{dy} 
& = \int d^2 \bm{b} \, \frac{\omega dN_\gamma(\omega,b)}{d\omega} \left| 
\int d\beta  d^2 \bm{r}\,
[\Psi_{V}^{*}\!\Psi_\gamma](\beta,r) 
\left(1 - \mathrm{e}^{-R_G\,\sigma_{q\bar q}(x,r)  T_A(b) /2 } 
\right)\right|^2 \\
& + (y \rightarrow -y)\,,
\label{eq:coherent-tot}
\end{align}
where the unresolved dipole kinematic variables -- the impact parameter $\bm{b}$ (with $b\equiv |\bm{b}|$) representing the transverse separation vector between the dipole and target nucleus centers-of-gravity, the \qq transverse separation $\bm{r}$ ($r\equiv |\bm{r}|$) and the photon momentum fraction carried by the quark $\beta$ -- are integrated out. In the $AA$ c.o.m.\ frame, with nucleon-nucleon ($NN$) energy of $\sqrt{s_{\rm NN}}$, the vector meson rapidity and the Bjorken variable are given by $y = \ln(2\omega/M_V)$ and $x = M_V\mathrm{e}^{-y}/\sqrt{s_{\rm NN}}$, respectively, with $\omega$ being the photon energy and $M_V$ -- the vector meson mass. In the following, each element of Eq.~(\ref{eq:coherent-tot}) will be described in detail.

The flux of Weisz$\ddot{{\rm a}}$cker-Williams (WW) photons $\omega dN_\gamma /d\omega$ emitted by the projectile nucleus and found at impact parameter $\bm{b}$ reads \cite{Guzey:2016piu}:
\begin{equation}
    \omega \frac{dN_\gamma(\omega,b)}{d\omega} = \int d^2 \bm{b}_\gamma \frac{\omega d^3N_{\gamma}(\omega,b_\gamma)}{d \omega d^2 \bm{b}_\gamma}
    \exp \left( -\sigma_\text{NN}^\text{tot} \int d^2 \bm{b}' T_A(b') T_A(|\bm{b}_{AA}-\bm{b}'|) \right) \, .
    \label{eq:photon_flux}
\end{equation}
Here, the exponential factor represents the probability of not having a strong interaction between the nuclei separated by impact parameter $\bm{b}_{AA} = \bm{b} - \bm{b}_\gamma$. In the argument of this exponential, the total $NN$ cross section $\sigma_{\rm NN}^{\text{tot}}$ \cite{COMPETE:2002jcr} accounts for strong interactions between the nucleons that would effectively occur at any impact parameter distance $b'$ from the center of the target nucleus. Besides, the standard WW method \cite{vonWeizsacker:1934nji,Williams:1934ad} considering the nucleus as a punctual charge provides the following quasi-real photon distribution at the $b_\gamma$ distance from the projectile nucleus center:
\begin{eqnarray}
\frac{\omega d^3N_\gamma(\omega,b_\gamma)}{d\omega d^2\bm{b}_\gamma} = 
\frac{Z^2\alpha_{\rm em} \omega^2}{\pi^2 \gamma^2} 
K^2_1\left(\frac{b_\gamma\, \omega}{\gamma}\right)
\,, 
\label{WW-flux}
\end{eqnarray}
where $\gamma = \sqrt{s_{\rm NN}}/2m_p$ is the Lorentz factor, with the proton mass given by $m_p=0.938$ GeV, and $Z$ is the charge of the projectile nucleus.

The nuclear thickness function $T_A(b)$ entering both Eqs.~(\ref{eq:coherent-tot}) and (\ref{eq:photon_flux}) is given by the integral of the nuclear density $\rho_A(b,z)$ over the longitudinal coordinate $z$ as
\begin{equation}
    T_A(b) = \int_{-\infty}^{+\infty}dz\,\rho_A(b,z)\,.
\end{equation} 
In this work, the Woods-Saxon parameterization \cite{Woods:1954zz} was chosen to represent the nuclear density,
\begin{equation}
\rho_A(b,z) = \frac{N_A}{1 + \exp\left[\left(\sqrt{b^2 + z^2} - R_C\right)/\delta\right]} \,, 
\end{equation}
where the prefactor $N_A$ is found from the normalization condition $\int d^2\bm{b}\,T_A(b) = A$, and the parameters for lead (Pb) nucleus $R_C = 6.62$ fm and $\delta = 0.546$ fm were taken from Ref.~\cite{Euteneuer:1978qw}.

At high energies, the \qq coherence length is considered to be much bigger than the nucleus radius ($l_c \gg R_A$). In this regime, the transverse dipole separation $r$ remains approximately constant, and the photon fluctuates into the \qq pair a long time before it passes through the nucleus. 

The splitting of a real transversely polarized photon into the \qq pair is described by the LF photon wave function ($\Psi_\gamma$). For small dipoles, it is known in perturbative QED since a long ago \cite{Kogut:1969xa, Bjorken:1970ah,Dosch:1996ss,Lepage:1980fj} (see also Ref.~\cite{Forshaw:2003ki}), and we do not reproduce it explicitly here.

On the other hand, the fact that the vector meson wave function ($\Psi_{V}$) describes the formation of a bound state prevents the use of perturbative methods to find its form. Instead, we chose the holographic approach developed by Brodsky and Téramond in Refs.~\cite{Brodsky:2008pg,Brodsky:2014yha}. It provides LF wave functions for all three light vector mesons that we are interested in, both in the ground and excited states. The holographic approach employs a semiclassical approximation, with finite quark mass $m_q$ corrections, which enables the following factorization for the scalar part of the vector meson wave function:
\begin{equation}
\psi(\beta, \zeta, \varphi) = e^{i L \varphi} \sqrt{\beta(1-\beta)}  \frac{\phi(\zeta)}{\sqrt{2 \pi \zeta}}\, 
\mathrm{e}^{\frac{m_q^2}{\beta}+\frac{m_q^2}{1-\beta}} \,,
\label{eq:factorized_wf_with_quark_mass}
\end{equation}
where $L$ is the orbital quantum momentum, $\varphi$ is the azimuthal angle, and $\zeta^2=\beta(1-\beta) r^2$. The function $\phi(\zeta)$ encodes the dynamics of the vector meson bound state and has the shape of harmonic-oscillator type wave functions, with $n$ being the principal quantum number,
\begin{equation}
\phi_{n,L}(\zeta) = \kappa^{1+L} \sqrt{\frac{2 n !}{(n+L) !}} \zeta^{1 / 2+L} \exp \left(\frac{-\kappa^2 \zeta^2}{2}\right) L_n^L\left(\kappa^2 \zeta^2\right) \,,
\label{eq:harmonic_oscilator_wave_equation}
\end{equation}
where $L_n^L\left(\kappa^2 \zeta^2\right)$ are the Laguerre polynomials. 

In Eq.~\ref{eq:harmonic_oscilator_wave_equation}, the $\kappa$ parameter represents the strength of the interaction between $q$ and $\bar q$. In our previous work~\cite{Henkels:2022bne} where we studied proton targets, we used a vector meson mass dependent $\kappa$ parameter, more specifically $\kappa \equiv M_{V(n=0)}/\sqrt{2}$. It provided a good description not only of $\rho$ and $\omega$ data but also of the available $\phi$ data, so we use the same $\kappa$ parameter in this present work.

In order to build $\Psi_V$ from $\psi$ given in Eq.~(\ref{eq:factorized_wf_with_quark_mass}), the vector meson polarization and quark helicity dependent structure of the same form as that for the photon $\gamma \to q\bar q$ LF wave function has been implemented (see e.g.~Ref.~\cite{Kowalski:2006hc} for further details). As a result, the overlap of the photon and vector meson LF wave functions for transverse $\gamma$, $V$ polarizations is given by
\begin{equation}
[\Psi_{V}^{*}\!\Psi_\gamma](\beta,r) =  \frac{e_f e \sqrt{N_c}}{(2\pi)^{3/2}2\beta(1-\beta)}\left[ m_f^2 K_0(\epsilon r)  \phi_{n, L}( \zeta) - \left(\beta^2 + (1-\beta)^2\right)\epsilon K_1(\epsilon r) \partial_r \phi_{n, L}(\zeta)\right] \,,
\end{equation}
where $K_{0,1}$ are the modified Bessel functions of the second kind, $\epsilon^2 = \beta (1-\beta)Q^2 + m_q^2$, with $Q^2=0$ in the case of photoproduction, $N_c = 3
$ is the number of colors in QCD, $e$ is the electromagnetic coupling constant, while the light quark flavor $f=u,d,s$ dependence enters through the values of the quark charge $e_f$ (in units of electron charge) and masses $m_{u,d} = 0.14$ GeV and $m_s = 0.35$ GeV.

The amplitude for the elastic scattering of the color dipole by a nucleon is proportional to the dipole cross section $\sigma_{q\bar q}(x,r)$. The Glauber--Gribov model is then employed to evaluate the dipole-nucleus amplitude \cite{Gribov:1968jf} at fixed impact parameter $b$, which results from the superposition of all possible successive elastic scatterings between the dipole and nucleons in the target nucleus. The result, for heavy nucleus, is the one minus exponential expression in Eq.~\ref{eq:coherent-tot}, that provides a suppression of the nuclear cross section compared to the free nucleon one.

In this work, we employ the GBW model for the dipole cross section, which relies on a simple saturated ansatz,
\begin{equation}
\sigma_{q\bar q}(x,r)=\sigma_{0}\left(1-e^{-r^2Q_s^2(x)/4}\right) \,,
\label{GBW_cross}
\end{equation}
where parameters were taken from the 2017 fit~\cite{Golec-Biernat:2017lfv} to HERA DIS data: the dipole cross section at large separations $\sigma_0=27.32$\,mb and the saturation scale $Q_{s}^{2}(x) = Q_{0}^{2}\left(x_{0}/x\right)^{\lambda}$, with $Q_{0}^{2}=1$\,GeV$^2$, $x_{0}=0.42 \times 10^{-4}$ and $\lambda=0.248$.

Some corrections are now in place. First, the dipole cross section is multiplied by the skewness factor $S_g$. This accounts for the fact that the exchanged gluons can carry different momentum fractions and is found as \cite{Shuvaev:1999ce}
\begin{equation}
S_g(\lambda) = \frac{2^{2\lambda + 3}}{\sqrt{\pi}}\frac{\Gamma(\lambda + 5/2)}{\Gamma (\lambda + 4)}\,,
\end{equation}
where $\lambda = \partial\ln\sigma_{q \bar q}/\partial\ln(1/x)$.

Second, we also included a contribution of the real part of the forward amplitude, calculated through the real-to-imaginary parts ratio of the photon-proton forward scattering amplitude \cite{Ivanov:2002kc}, via a replacement
\begin{equation}
\sigma_{q\bar{q}}(x,r) \Rightarrow \sigma_{q\bar{q}}(x,r)\left(1 - i \frac{\pi}{2}\frac{\partial\ln\sigma_{q\bar{q}}}{\partial\ln(1/x)}\right) \,. \label{sigdip_skew}
\end{equation}

At last, a crucial nuclear effect present in typical high-energy reactions off nuclear targets is, as named in the parton model, the gluon shadowing, that is another source of suppression of the nuclear cross section compared to the free nucleon one. In the color dipole approach, this effect is a result of the fact that the photon wave function fluctuates also into the higher Fock states containing gluons, i.e., $\left| q\bar{q} G \right \rangle$, $\left| q\bar{q} 2G \right \rangle$, ... , $\left| q\bar{q} n G \right \rangle$, that have a smaller lifetime than the lowest Fock $\left| q\bar{q} \right \rangle$ state~\cite{Kopeliovich:2001xj,Kopeliovich:2020has}. Such states are considered frozen in the dipole-proton interaction as their coherence length is greater than the proton radius $l_c^{q\Bar{q}g}, l_c^{q\Bar{q}gg}, \cdots  \gg R_p$. As such, the dipole-proton cross section parameterizations are fitted to data where all of these higher Fock states contribute. This is not true for reactions off nuclear targets where one has to take into consideration that the higher Fock states have a smaller coherence length than the nuclear radius and as such will have smaller contributions. 

An effective way to incorporate such a gluon shadowing effect into the present calculation and additionally suppress the nuclear cross section is by ``renormalizing'' the dipole cross section with an additional $x$- and scale-dependent suppression factor $R_G$ as
\begin{equation}
\sigma_{q\bar q}(x,r) \rightarrow  \sigma_{q\bar{q}}(x,r) R_G(x,\mu^2) \, .
\end{equation}
In a previous work of Ref.~\cite{Henkels:2020kju} on heavy vector meson photoproduction, we evaluated this factor as a ratio between the gluon density of a nucleon inside a given nucleus and the one found in a free nucleon using a standard set of nucleon and nuclear PDFs at the scale of $\mu^2 = M_V^2/4$. However, in the considered case of light vector meson photoproduction, the absence of a sufficiently hard scale prevents the use of this method, since the existing PDF parameterizations are justified in the perturbative regime and evolved at scales larger than the charm quark mass $m_c = 1.3$ GeV. 

Instead, here we adopt an alternative approach by extracting an optimal value for $R_G$ from a fit to the available $\rho(1S)$ and $F_2$ at low $Q^2$ data, as will be discussed in the next section. In this way, any nuclear contribution (eikonal or sub-eikonal) other than the Glauber--Gribov formalism will be included in the $R_G$ factor. We take the viewpoint that other corrections that are also present in the proton target case are effectively absorbed into the $\sigma_{q\bar{q}}$ and the quark masses.

%-------------------
\section{Results}
\label{results}
%-------------------

\begin{figure}[!h]
    \centering
    \includegraphics[width = .6\textwidth]{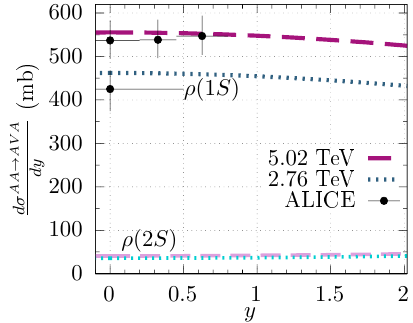}
    \caption{The differential cross section for coherent $\rho(1S, 2S)$ photoproduction in PbPb collisions as a function of meson rapidity $y$, calculated using the holographic wave function in conjunction with the GBW dipole model. The gluon shadowing effect, parameterized by $R_G = 0.85$, was fitted to the data as described in the text.}
    \label{fig:rhofig}
\end{figure}

The formalism discussed above enables us to perform an analysis of the coherent photoproduction of light mesons in the dipole picture, with a heavy nucleus as a target. Fig.~\ref{fig:rhofig} shows the total nuclear cross section of $\rho$ meson photoproduction as a function of the meson rapidity. As mentioned briefly above and studied in previous works of e.g.~Refs.~\cite{Henkels:2020kju, Henkels:2022bne, Kopeliovich:2022jwe}, the inclusion of the gluon shadowing effect leads to a suppression of the total cross section required to describe the available data for the photoproduction of heavy states such as $J/\Psi$ off heavy nuclei \cite{Henkels:2020kju}. We pick the rapidity range of $|y|<2$, this choice  ensures that the Björken-$x$ stays at a sufficiently small value (from $\sim 10^{-5}$ up to $\sim 10^{-3}$ at ALICE $2.76$ TeV), where the dipole formalism is justified.

In this work focusing on light vector mesons, we adopted an effective value of the suppression factor $R_G = 0.85$ without impact parameter dependence. Such a value has been obtained from a fit to the four available data points of $\rho$ photoproduction at ALICE \cite{ALICE:2015nbw, ALICE:2020ugp} (with $x \approx 10^{-4}$ and $M_\rho^2/4 \approx 0.15$ GeV$^2$) and the lowest-scale $Q^2$ data point $F_2^A/(AF_2^p) = 0.628 \pm 0.048 \pm 0.079$ for the ratio of the nuclear to nucleon structure function measured by the E665 collaboration \cite{E665:1995xur} (with $x_{\rm Bj}\subset (1-3.7)\cdot 10^{-4}$ and $\langle Q^2 \rangle = 0.15$ GeV$^2$). For $F^A_2$, we follow the approach of Refs.~\cite{Armesto:2002ny, Krelina:2020ipn, Kopeliovich:1998gv} that rests on the same physics that are used in our vector meson production case. Our calculation leads to $F_2^A/(AF_2^p) = 0.537$ and the fitted value of $R_G$ maintained all data description within error bars. 

To judge if the fitted $R_G$ indeed captures nuclear effects, we compare our predictions for the proton case with experimental data from the CMS collaboration. Specifically, for the $\gamma p \rightarrow \rho p$ process at $W_{\gamma p} = 59.2$ GeV, the CMS data reports a cross section of $\sigma_{\gamma p}^{\text{exp}} = 9.9 \pm 1.84$ $\mu$b~\cite{CMS:2019awk}. This energy is close to the $W_{\gamma p} = 62$ GeV in our nuclear case at midrapidity $y = 0$. Within our model, we calculate $\sigma_{\gamma p} = 9.15$ $\mu$b, which lies within the error bars of the experimental value, indicating good agreement between our theoretical predictions and the data.

We remark that $R_G$ is expected to depend on both $x$ and $\mu$. However, the data we have considered show similar values for these variables, which motivates us to approximate $R_G$ by a single value for the five points. The shadowing effect certainly depends on $x$, for this reason, we choose not to explore larger rapidities in our analysis. Furthermore, should any experiment succeed in extracting the $\gamma A$ cross section, direct comparisons can be made with the $\gamma A$ figure presented below. For instance, measurements at LHCb rapidities would be particularly valuable for determining this dependence of $R_G$ on $x$.

%%%%%%%%%%%%%%%%%%%%%%%%%%%%%%%%%%%%%%%%%%%%%%
%%%%%%%%%%%%%%%%%%%%%%%%%%%%%%%%%%%%%%%%%%%%%%
\begin{figure}[!h]
    \centering
    \includegraphics[width = .6\textwidth]{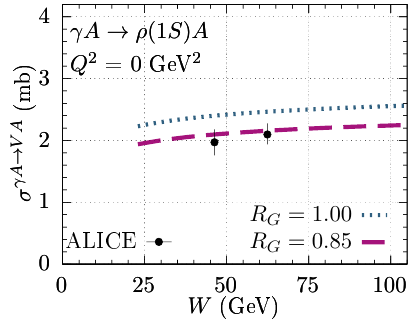}
    \caption{Photoproduction of $\rho(1S)$ in $\gamma$Pb collisions as a function of center-of-mass energy $W$. Our results are compared to the ALICE data extracted in Refs. \cite{Frankfurt:2015cwa, ALICE:2021jnv}. We show predictions with the adjusted gluon shadowing effect ($R_G = 0.85$) and without it ($R_G = 1$).}
    \label{fig:rhofig_gammaA}
\end{figure}
In Fig.~\ref{fig:rhofig}, we show the results for the differential (in rapidity) $\rho(1S)$ photoproduction cross section compared to the available ALICE~\cite{ALICE:2015nbw, ALICE:2020ugp} data at $\sqrt{s_{\rm NN}} = 2.76$ TeV and $5.02$ TeV for PbPb UPCs. We notice that the dipole model description, based upon the simplest GBW parameterization for the universal dipole cross section, the holographic LF meson wave functions and the phenomenological value for the gluon shadowing suppression factor $R_G = 0.85$ as described above, is capable of describing all four available data points. 

In Fig.~\ref{fig:rhofig_gammaA}, we show the photoproduction of $\rho(1S)$ in $\gamma$Pb collisions as a function of center-of-mass energy $W$. Our results are compared to the ALICE data extracted in Refs. \cite{Frankfurt:2015cwa, ALICE:2021jnv}. We show predictions with the adjusted gluon shadowing effect ($R_G = 0.85$) and without it ($R_G = 1$).

\begin{figure}[!h]
    \centering
    \includegraphics[width = .6\textwidth]{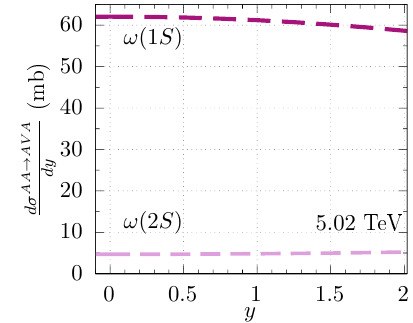}
    \caption{Predictions for the differential cross section of coherent $\omega(1S, 2S)$ photoproduction in PbPb UPCs as a function of the meson rapidity $y$. The method is the same as in Fig.~\ref{fig:rhofig}.}
    \label{fig:omegafig}
\end{figure}

Considering how close the masses of the $\omega(1S)$ and $\rho(1S)$ mesons are, we show predictions in Fig.~\ref{fig:omegafig} for the total nuclear cross section of $\omega$-meson photoproduction calculated with the same $R_G$ value. Future measurements of $\omega(1S)$ and $\rho(1S)$ nuclear photoproduction would tell us whether the gluon shadowing should be different in the two cases. We also make predictions for the excited $\rho(2S)$ and $\omega(2S)$ states with the extrapolation of using the same gluon shadowing.

%%%%%%%%%%%%%%%%%%%%%%%%%%%%%%%%%%%%%%%%%%%%%%%%%%%%%%%%%%%%%%%%%%%%%%%%
%%%%%%%%%%%%%%%%%%%%%%%%%%%%%%%%%%%%%%%%%%%%%%%%%%%%%%%%%%%%%%%%%%%%%%%%

\begin{figure}[!h]
    \centering
    \includegraphics[width = .6\textwidth]{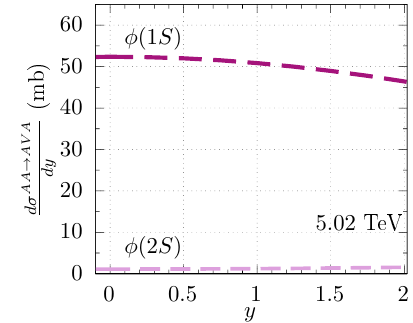}
    \caption{Predictions for the differential cross section of coherent $\phi(1S, 2S)$ photoproduction in PbPb UPCs as a function of the meson rapidity $y$. The method is the same as in Fig.~\ref{fig:rhofig}.}
    \label{fig:phifig}
\end{figure}

We also make predictions for the photoproduction cross section of $\phi$ meson states. Fig.~\ref{fig:phifig} shows the total cross section for the differential $\phi(1S,2S)$ photoproduction cross section as a function of the meson rapidity $y$. It is worth mentioning that measurements of the $\phi$ nuclear photoproduction process could also help us in the characterization of the gluon shadowing effect at a somewhat larger scale than the one considered for $\rho$ and $\omega$ mesons. Since there is no such available data yet, we make these predictions implementing the same $R_G = 0.85$ suppression factor as a suitable reference point.
%%%%%%%%%%%%%%%%%%%%%%%%%%%%%%%%%%%%%%%%%%%%%%%%%%%%%%%%%%%%%
%%%%%%%%%%%%%%%%%%%%%%%%%%%%%%%%%%%%%%%%%%%%%%%%%%%%%%%%%%%%%

\begin{figure}[!h]
    \centering
    \includegraphics[width = .6\textwidth]{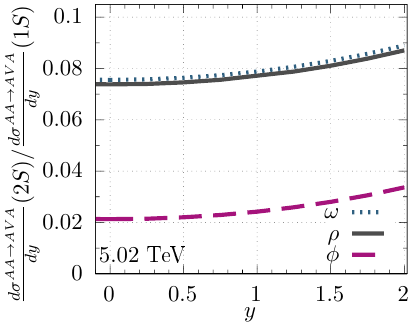}
    \caption{Predictions for the ratio between the excited-state and the ground-state differential cross sections as a function of rapidity, for three light vector mesons. The method is the same as in Fig.~\ref{fig:rhofig}.}
    \label{fig:excfig}
\end{figure}

Finally, in Fig.~\ref{fig:excfig} we show predictions for the ratio between the excited-state and the ground-state differential cross sections as a function of rapidity, for three light vector mesons $\omega$, $\rho$ and $\phi$. We notice that such a ratio increases with rapidity similarly for each of the three mesons. At forward rapidities, the dominant contribution to the cross section comes from the kinematical configurations with Bjorken $x$ is larger than the longitudinal momentum fraction of the photon, i.e.~the $(y \rightarrow - y )$ contribution in Eq.~(\ref{eq:coherent-tot}). Indeed, for larger $x$, the larger $r$ color dipoles dominate the cross section, such that the production of the ground state (with no node in the wave function) would be more suppressed in comparison with the production of the wider excited states (with nodes).

%%%%%%%%%%%%%%%%%%%%%%%%%%%%%%%%%%%%%%%%%%%%%%%%%%%%%%%%%%%%%%%%%%%%%
%%%%%%%%%%%%%%%%%%%%%%%%%%%%%%%%%%%%%%%%%%%%%%%%%%%%%%%%%%%%%%%%%%%%%

We believe that future measurements of exclusive $\omega$, $\rho$ and $\phi$ photoproduction in $AA$ UPCs could further constrain the dipole approach with its main ingredients -- the universal dipole cross section, the gluon shadowing corrections and the LF meson wave functions for light excited states -- all being the sources of theoretical uncertainties in the present analysis. 

%----------------------------------------------------------
\section{Conclusion}
\label{conclusion}
%----------------------------------------------------------

We have analyzed coherent photoproduction of light vector mesons $\omega$, $\rho$ and $\phi$ in $AA$ UPCs. Despite being an essentially soft process, the color dipole approach is expected to be a suitable framework for its studies due to universality of the dipole description which does not rely on QCD factorization. In order to study nuclear targets in this work, we employ the Glauber--Gribov formalism, that naturally implements a nuclear suppression of the dipole scattering amplitude that is sometimes called ``quark shadowing''. 

One additional important ingredient of the dipole propagation in the nuclear environment is the gluon shadowing effect due to the presence of the higher Fock states in the photon wave function. At larger scales typical e.g.~for heavy vector meson photoproduction such as $J/\psi$, such a contribution is reasonably well understood and can be either phenomenologically extracted from the experimental data on nuclear PDFs in the parton model or determined in the Green functions formalism for a dipole propagating through the nuclear medium. However, at lower scales such as those involved in the current consideration of the light vector mesons, the aforementioned approaches are not well justified. 

In the absence of a theory for the gluon shadowing at lower scales, we have adopted a naive phenomenological description by introducing an overall multiplicative constant factor $R_G$ in the elementary dipole-target scattering amplitude. This factor has been extracted by a simple fit to the existing data at very small $Q^2$ and $x$ on the DIS structure function $F_2^A$ from the E665 Collaboration and on $\rho$ meson photoproduction from the LHC. This approach provides a good description of all five available data points yielding $R_G = 0.85$. 

Using the same suppression factor as an extrapolation to the soft regime, we have computed the coherent $\rho(2S)$, $\omega(1S,2S)$ and $\phi(1S,2S)$ photoproduction cross sections using the simplest GBW dipole model and the holographic model for corresponding vector meson wave functions. To reduce potentially large theoretical uncertainties, we have made predictions for the ratios between the excited-state and the ground-state photoproduction cross sections for $\omega$, $\rho$ and $\phi$ mesons. 

It is a persistent challenge to consistently describe the photoproduction observables for both the ground and excited states in the framework of a single universal approach. We believe that future measurements of light vector-meson photoproduction processes at the LHC and at the Electron-Ion Collider (EIC) and, in particular, the excited-to-ground states' ratios of the cross sections we predict in this work, would verify our current understanding of the underlined soft/nonperturbative QCD dynamics and help establish the correct production mechanism incorporating both the dipole-nucleus cross section (with an appropriate treatment of nuclear effects such as gluon shadowing), the Glauber--Gribov approach and the light vector-meson wave functions.
 
%--------------------------
\section*{Acknowledgments}
%--------------------------

This work was supported by Fapesc, INCT-FNA (464898/2014-5), and CNPq (Brazil) for CH, EGdO, and HT. This study was financed in part by the Coordenação de Aperfeiçoamento de Pessoal de Nível Superior -- Brasil (CAPES) -- Finance Code 001. R.P.~is supported in part by the Swedish Research Council grants, contract numbers 621-2013-4287 and 2016-05996, as well as by the European Research Council (ERC) under the European Union's Horizon 2020 research and innovation programme (grant agreement No 668679).

\bibliography{bib}

\begin{thebibliography}{65}%
\makeatletter
\providecommand \@ifxundefined [1]{%
 \@ifx{#1\undefined}
}%
\providecommand \@ifnum [1]{%
 \ifnum #1\expandafter \@firstoftwo
 \else \expandafter \@secondoftwo
 \fi
}%
\providecommand \@ifx [1]{%
 \ifx #1\expandafter \@firstoftwo
 \else \expandafter \@secondoftwo
 \fi
}%
\providecommand \natexlab [1]{#1}%
\providecommand \enquote  [1]{``#1''}%
\providecommand \bibnamefont  [1]{#1}%
\providecommand \bibfnamefont [1]{#1}%
\providecommand \citenamefont [1]{#1}%
\providecommand \href@noop [0]{\@secondoftwo}%
\providecommand \href [0]{\begingroup \@sanitize@url \@href}%
\providecommand \@href[1]{\@@startlink{#1}\@@href}%
\providecommand \@@href[1]{\endgroup#1\@@endlink}%
\providecommand \@sanitize@url [0]{\catcode `\\12\catcode `\$12\catcode
  `\&12\catcode `\#12\catcode `\^12\catcode `\_12\catcode `\%12\relax}%
\providecommand \@@startlink[1]{}%
\providecommand \@@endlink[0]{}%
\providecommand \url  [0]{\begingroup\@sanitize@url \@url }%
\providecommand \@url [1]{\endgroup\@href {#1}{\urlprefix }}%
\providecommand \urlprefix  [0]{URL }%
\providecommand \Eprint [0]{\href }%
\providecommand \doibase [0]{http://dx.doi.org/}%
\providecommand \selectlanguage [0]{\@gobble}%
\providecommand \bibinfo  [0]{\@secondoftwo}%
\providecommand \bibfield  [0]{\@secondoftwo}%
\providecommand \translation [1]{[#1]}%
\providecommand \BibitemOpen [0]{}%
\providecommand \bibitemStop [0]{}%
\providecommand \bibitemNoStop [0]{.\EOS\space}%
\providecommand \EOS [0]{\spacefactor3000\relax}%
\providecommand \BibitemShut  [1]{\csname bibitem#1\endcsname}%
\let\auto@bib@innerbib\@empty
%</preamble>
\bibitem [{\citenamefont {Adler}\ \emph {et~al.}(2002)\citenamefont {Adler}
  \emph {et~al.}}]{STAR:2002caw}%
  \BibitemOpen
  \bibfield  {author} {\bibinfo {author} {\bibfnamefont {C.}~\bibnamefont
  {Adler}} \emph {et~al.} (\bibinfo {collaboration} {STAR}),\ }\href {\doibase
  10.1103/PhysRevLett.89.272302} {\bibfield  {journal} {\bibinfo  {journal}
  {Phys. Rev. Lett.}\ }\textbf {\bibinfo {volume} {89}},\ \bibinfo {pages}
  {272302} (\bibinfo {year} {2002})},\ \Eprint
  {http://arxiv.org/abs/nucl-ex/0206004} {arXiv:nucl-ex/0206004} \BibitemShut
  {NoStop}%
\bibitem [{\citenamefont {Agakishiev}\ \emph {et~al.}(2012)\citenamefont
  {Agakishiev} \emph {et~al.}}]{STAR:2011wtm}%
  \BibitemOpen
  \bibfield  {author} {\bibinfo {author} {\bibfnamefont {G.}~\bibnamefont
  {Agakishiev}} \emph {et~al.} (\bibinfo {collaboration} {STAR}),\ }\href
  {\doibase 10.1103/PhysRevC.85.014910} {\bibfield  {journal} {\bibinfo
  {journal} {Phys. Rev. C}\ }\textbf {\bibinfo {volume} {85}},\ \bibinfo
  {pages} {014910} (\bibinfo {year} {2012})},\ \Eprint
  {http://arxiv.org/abs/1107.4630} {arXiv:1107.4630 [nucl-ex]} \BibitemShut
  {NoStop}%
\bibitem [{\citenamefont {Afanasiev}\ \emph {et~al.}(2009)\citenamefont
  {Afanasiev} \emph {et~al.}}]{PHENIX:2009xtn}%
  \BibitemOpen
  \bibfield  {author} {\bibinfo {author} {\bibfnamefont {S.}~\bibnamefont
  {Afanasiev}} \emph {et~al.} (\bibinfo {collaboration} {PHENIX}),\ }\href
  {\doibase 10.1016/j.physletb.2009.07.061} {\bibfield  {journal} {\bibinfo
  {journal} {Phys. Lett. B}\ }\textbf {\bibinfo {volume} {679}},\ \bibinfo
  {pages} {321} (\bibinfo {year} {2009})},\ \Eprint
  {http://arxiv.org/abs/0903.2041} {arXiv:0903.2041 [nucl-ex]} \BibitemShut
  {NoStop}%
\bibitem [{\citenamefont {Khachatryan}\ \emph {et~al.}(2017)\citenamefont
  {Khachatryan} \emph {et~al.}}]{CMS:2016itn}%
  \BibitemOpen
  \bibfield  {author} {\bibinfo {author} {\bibfnamefont {V.}~\bibnamefont
  {Khachatryan}} \emph {et~al.} (\bibinfo {collaboration} {CMS}),\ }\href
  {\doibase 10.1016/j.physletb.2017.07.001} {\bibfield  {journal} {\bibinfo
  {journal} {Phys. Lett. B}\ }\textbf {\bibinfo {volume} {772}},\ \bibinfo
  {pages} {489} (\bibinfo {year} {2017})},\ \Eprint
  {http://arxiv.org/abs/1605.06966} {arXiv:1605.06966 [nucl-ex]} \BibitemShut
  {NoStop}%
\bibitem [{\citenamefont {Abelev}\ \emph {et~al.}(2013)\citenamefont {Abelev}
  \emph {et~al.}}]{ALICE:2012yye}%
  \BibitemOpen
  \bibfield  {author} {\bibinfo {author} {\bibfnamefont {B.}~\bibnamefont
  {Abelev}} \emph {et~al.} (\bibinfo {collaboration} {ALICE}),\ }\href
  {\doibase 10.1016/j.physletb.2012.11.059} {\bibfield  {journal} {\bibinfo
  {journal} {Phys. Lett. B}\ }\textbf {\bibinfo {volume} {718}},\ \bibinfo
  {pages} {1273} (\bibinfo {year} {2013})},\ \Eprint
  {http://arxiv.org/abs/1209.3715} {arXiv:1209.3715 [nucl-ex]} \BibitemShut
  {NoStop}%
\bibitem [{\citenamefont {Abbas}\ \emph {et~al.}(2013)\citenamefont {Abbas}
  \emph {et~al.}}]{ALICE:2013wjo}%
  \BibitemOpen
  \bibfield  {author} {\bibinfo {author} {\bibfnamefont {E.}~\bibnamefont
  {Abbas}} \emph {et~al.} (\bibinfo {collaboration} {ALICE}),\ }\href {\doibase
  10.1140/epjc/s10052-013-2617-1} {\bibfield  {journal} {\bibinfo  {journal}
  {Eur. Phys. J. C}\ }\textbf {\bibinfo {volume} {73}},\ \bibinfo {pages}
  {2617} (\bibinfo {year} {2013})},\ \Eprint {http://arxiv.org/abs/1305.1467}
  {arXiv:1305.1467 [nucl-ex]} \BibitemShut {NoStop}%
\bibitem [{\citenamefont {Adam}\ \emph
  {et~al.}(2015{\natexlab{a}})\citenamefont {Adam} \emph
  {et~al.}}]{ALICE:2015nmy}%
  \BibitemOpen
  \bibfield  {author} {\bibinfo {author} {\bibfnamefont {J.}~\bibnamefont
  {Adam}} \emph {et~al.} (\bibinfo {collaboration} {ALICE}),\ }\href {\doibase
  10.1016/j.physletb.2015.10.040} {\bibfield  {journal} {\bibinfo  {journal}
  {Phys. Lett. B}\ }\textbf {\bibinfo {volume} {751}},\ \bibinfo {pages} {358}
  (\bibinfo {year} {2015}{\natexlab{a}})},\ \Eprint
  {http://arxiv.org/abs/1508.05076} {arXiv:1508.05076 [nucl-ex]} \BibitemShut
  {NoStop}%
\bibitem [{\citenamefont {Acharya}\ \emph {et~al.}()\citenamefont {Acharya}
  \emph {et~al.}}]{ALICE:2023svb}%
  \BibitemOpen
  \bibfield  {author} {\bibinfo {author} {\bibfnamefont {S.}~\bibnamefont
  {Acharya}} \emph {et~al.} (\bibinfo {collaboration} {ALICE}),\ }\href@noop {}
  {\ }\Eprint {http://arxiv.org/abs/2304.10928} {arXiv:2304.10928 [nucl-ex]}
  \BibitemShut {NoStop}%
\bibitem [{\citenamefont {Adam}\ \emph
  {et~al.}(2015{\natexlab{b}})\citenamefont {Adam} \emph
  {et~al.}}]{ALICE:2015nbw}%
  \BibitemOpen
  \bibfield  {author} {\bibinfo {author} {\bibfnamefont {J.}~\bibnamefont
  {Adam}} \emph {et~al.} (\bibinfo {collaboration} {ALICE}),\ }\href {\doibase
  10.1007/JHEP09(2015)095} {\bibfield  {journal} {\bibinfo  {journal} {JHEP}\
  }\textbf {\bibinfo {volume} {09}},\ \bibinfo {pages} {095} (\bibinfo {year}
  {2015}{\natexlab{b}})},\ \Eprint {http://arxiv.org/abs/1503.09177}
  {arXiv:1503.09177 [nucl-ex]} \BibitemShut {NoStop}%
\bibitem [{\citenamefont {Acharya}\ \emph {et~al.}(2020)\citenamefont {Acharya}
  \emph {et~al.}}]{ALICE:2020ugp}%
  \BibitemOpen
  \bibfield  {author} {\bibinfo {author} {\bibfnamefont {S.}~\bibnamefont
  {Acharya}} \emph {et~al.} (\bibinfo {collaboration} {ALICE}),\ }\href
  {\doibase 10.1007/JHEP06(2020)035} {\bibfield  {journal} {\bibinfo  {journal}
  {JHEP}\ }\textbf {\bibinfo {volume} {06}},\ \bibinfo {pages} {035} (\bibinfo
  {year} {2020})},\ \Eprint {http://arxiv.org/abs/2002.10897} {arXiv:2002.10897
  [nucl-ex]} \BibitemShut {NoStop}%
\bibitem [{\citenamefont {Jenkovszky}\ \emph {et~al.}(2022)\citenamefont
  {Jenkovszky}, \citenamefont {Rocha},\ and\ \citenamefont
  {Machado}}]{Jenkovszky:2022qnc}%
  \BibitemOpen
  \bibfield  {author} {\bibinfo {author} {\bibfnamefont {L.}~\bibnamefont
  {Jenkovszky}}, \bibinfo {author} {\bibfnamefont {E.~d.~S.}\ \bibnamefont
  {Rocha}}, \ and\ \bibinfo {author} {\bibfnamefont {M.~V.~T.}\ \bibnamefont
  {Machado}},\ }\href {\doibase 10.1016/j.physletb.2022.137585} {\bibfield
  {journal} {\bibinfo  {journal} {Phys. Lett. B}\ }\textbf {\bibinfo {volume}
  {835}},\ \bibinfo {pages} {137585} (\bibinfo {year} {2022})},\ \Eprint
  {http://arxiv.org/abs/2210.15749} {arXiv:2210.15749 [hep-ph]} \BibitemShut
  {NoStop}%
\bibitem [{\citenamefont {Nikolaev}\ and\ \citenamefont
  {Zakharov}(1994)}]{Nikolaev:1994kk}%
  \BibitemOpen
  \bibfield  {author} {\bibinfo {author} {\bibfnamefont {N.~N.}\ \bibnamefont
  {Nikolaev}}\ and\ \bibinfo {author} {\bibfnamefont {B.~G.}\ \bibnamefont
  {Zakharov}},\ }\href@noop {} {\bibfield  {journal} {\bibinfo  {journal} {J.
  Exp. Theor. Phys.}\ }\textbf {\bibinfo {volume} {78}},\ \bibinfo {pages}
  {598} (\bibinfo {year} {1994})},\ \bibinfo {note} {[Zh. Eksp. Teor.
  Fiz.105,1117(1994)]}\BibitemShut {NoStop}%
%%CITATION = JTPHE,78,598;%%
\bibitem [{\citenamefont {Mueller}\ and\ \citenamefont
  {Patel}(1994)}]{Mueller:1994jq}%
  \BibitemOpen
  \bibfield  {author} {\bibinfo {author} {\bibfnamefont {A.~H.}\ \bibnamefont
  {Mueller}}\ and\ \bibinfo {author} {\bibfnamefont {B.}~\bibnamefont
  {Patel}},\ }\href {\doibase 10.1016/0550-3213(94)90284-4} {\bibfield
  {journal} {\bibinfo  {journal} {Nucl. Phys. B}\ }\textbf {\bibinfo {volume}
  {425}},\ \bibinfo {pages} {471} (\bibinfo {year} {1994})},\ \Eprint
  {http://arxiv.org/abs/hep-ph/9403256} {arXiv:hep-ph/9403256} \BibitemShut
  {NoStop}%
\bibitem [{\citenamefont {Henkels}\ \emph {et~al.}(2023)\citenamefont
  {Henkels}, \citenamefont {de~Oliveira}, \citenamefont {Pasechnik},\ and\
  \citenamefont {Trebien}}]{Henkels:2022bne}%
  \BibitemOpen
  \bibfield  {author} {\bibinfo {author} {\bibfnamefont {C.}~\bibnamefont
  {Henkels}}, \bibinfo {author} {\bibfnamefont {E.~G.}\ \bibnamefont
  {de~Oliveira}}, \bibinfo {author} {\bibfnamefont {R.}~\bibnamefont
  {Pasechnik}}, \ and\ \bibinfo {author} {\bibfnamefont {H.}~\bibnamefont
  {Trebien}},\ }\href {\doibase 10.1140/epjc/s10052-023-11706-5} {\bibfield
  {journal} {\bibinfo  {journal} {Eur. Phys. J. C}\ }\textbf {\bibinfo {volume}
  {83}},\ \bibinfo {pages} {551} (\bibinfo {year} {2023})},\ \Eprint
  {http://arxiv.org/abs/2207.13756} {arXiv:2207.13756 [hep-ph]} \BibitemShut
  {NoStop}%
\bibitem [{\citenamefont {Nikolaev}\ and\ \citenamefont
  {Zakharov}(1991)}]{Nikolaev:1990ja}%
  \BibitemOpen
  \bibfield  {author} {\bibinfo {author} {\bibfnamefont {N.~N.}\ \bibnamefont
  {Nikolaev}}\ and\ \bibinfo {author} {\bibfnamefont {B.~G.}\ \bibnamefont
  {Zakharov}},\ }\href {\doibase 10.1007/BF01483577} {\bibfield  {journal}
  {\bibinfo  {journal} {Z. Phys. C}\ }\textbf {\bibinfo {volume} {49}},\
  \bibinfo {pages} {607} (\bibinfo {year} {1991})}\BibitemShut {NoStop}%
\bibitem [{\citenamefont {Nikolaev}\ and\ \citenamefont
  {Zakharov}(1992)}]{Nikolaev:1991et}%
  \BibitemOpen
  \bibfield  {author} {\bibinfo {author} {\bibfnamefont {N.}~\bibnamefont
  {Nikolaev}}\ and\ \bibinfo {author} {\bibfnamefont {B.~G.}\ \bibnamefont
  {Zakharov}},\ }\href {\doibase 10.1007/BF01597573} {\bibfield  {journal}
  {\bibinfo  {journal} {Z. Phys. C}\ }\textbf {\bibinfo {volume} {53}},\
  \bibinfo {pages} {331} (\bibinfo {year} {1992})}\BibitemShut {NoStop}%
\bibitem [{\citenamefont {Henkels}\ \emph {et~al.}(2021)\citenamefont
  {Henkels}, \citenamefont {de~Oliveira}, \citenamefont {Pasechnik},\ and\
  \citenamefont {Trebien}}]{Henkels:2020qvo}%
  \BibitemOpen
  \bibfield  {author} {\bibinfo {author} {\bibfnamefont {C.}~\bibnamefont
  {Henkels}}, \bibinfo {author} {\bibfnamefont {E.~G.}\ \bibnamefont
  {de~Oliveira}}, \bibinfo {author} {\bibfnamefont {R.}~\bibnamefont
  {Pasechnik}}, \ and\ \bibinfo {author} {\bibfnamefont {H.}~\bibnamefont
  {Trebien}},\ }\href {\doibase 10.1103/PhysRevD.104.054008} {\bibfield
  {journal} {\bibinfo  {journal} {Phys. Rev. D}\ }\textbf {\bibinfo {volume}
  {104}},\ \bibinfo {pages} {054008} (\bibinfo {year} {2021})},\ \Eprint
  {http://arxiv.org/abs/2009.14158} {arXiv:2009.14158 [hep-ph]} \BibitemShut
  {NoStop}%
\bibitem [{\citenamefont {Gon\c{c}alves}\ \emph {et~al.}(2022)\citenamefont
  {Gon\c{c}alves}, \citenamefont {Martins},\ and\ \citenamefont
  {Sena}}]{Goncalves:2022wzq}%
  \BibitemOpen
  \bibfield  {author} {\bibinfo {author} {\bibfnamefont {V.~P.}\ \bibnamefont
  {Gon\c{c}alves}}, \bibinfo {author} {\bibfnamefont {D.~E.}\ \bibnamefont
  {Martins}}, \ and\ \bibinfo {author} {\bibfnamefont {C.~R.}\ \bibnamefont
  {Sena}},\ }\href {\doibase 10.1140/epja/s10050-022-00664-3} {\bibfield
  {journal} {\bibinfo  {journal} {Eur. Phys. J. A}\ }\textbf {\bibinfo {volume}
  {58}},\ \bibinfo {pages} {18} (\bibinfo {year} {2022})}\BibitemShut {NoStop}%
\bibitem [{\citenamefont {Cepila}\ \emph {et~al.}(2019)\citenamefont {Cepila},
  \citenamefont {Nemchik}, \citenamefont {Krelina},\ and\ \citenamefont
  {Pasechnik}}]{Cepila:2019skb}%
  \BibitemOpen
  \bibfield  {author} {\bibinfo {author} {\bibfnamefont {J.}~\bibnamefont
  {Cepila}}, \bibinfo {author} {\bibfnamefont {J.}~\bibnamefont {Nemchik}},
  \bibinfo {author} {\bibfnamefont {M.}~\bibnamefont {Krelina}}, \ and\
  \bibinfo {author} {\bibfnamefont {R.}~\bibnamefont {Pasechnik}},\ }\href
  {\doibase 10.1140/epjc/s10052-019-7016-9} {\bibfield  {journal} {\bibinfo
  {journal} {Eur. Phys. J. C}\ }\textbf {\bibinfo {volume} {79}},\ \bibinfo
  {pages} {495} (\bibinfo {year} {2019})},\ \Eprint
  {http://arxiv.org/abs/1901.02664} {arXiv:1901.02664 [hep-ph]} \BibitemShut
  {NoStop}%
\bibitem [{\citenamefont {Kopeliovich}\ \emph {et~al.}(2002)\citenamefont
  {Kopeliovich}, \citenamefont {Nemchik}, \citenamefont {Schafer},\ and\
  \citenamefont {Tarasov}}]{Kopeliovich:2001xj}%
  \BibitemOpen
  \bibfield  {author} {\bibinfo {author} {\bibfnamefont {B.~Z.}\ \bibnamefont
  {Kopeliovich}}, \bibinfo {author} {\bibfnamefont {J.}~\bibnamefont
  {Nemchik}}, \bibinfo {author} {\bibfnamefont {A.}~\bibnamefont {Schafer}}, \
  and\ \bibinfo {author} {\bibfnamefont {A.~V.}\ \bibnamefont {Tarasov}},\
  }\href {\doibase 10.1103/PhysRevC.65.035201} {\bibfield  {journal} {\bibinfo
  {journal} {Phys. Rev. C}\ }\textbf {\bibinfo {volume} {65}},\ \bibinfo
  {pages} {035201} (\bibinfo {year} {2002})},\ \Eprint
  {http://arxiv.org/abs/hep-ph/0107227} {arXiv:hep-ph/0107227} \BibitemShut
  {NoStop}%
\bibitem [{\citenamefont {Golec-Biernat}\ and\ \citenamefont
  {Sapeta}(2018)}]{Golec-Biernat:2017lfv}%
  \BibitemOpen
  \bibfield  {author} {\bibinfo {author} {\bibfnamefont {K.}~\bibnamefont
  {Golec-Biernat}}\ and\ \bibinfo {author} {\bibfnamefont {S.}~\bibnamefont
  {Sapeta}},\ }\href {\doibase 10.1007/JHEP03(2018)102} {\bibfield  {journal}
  {\bibinfo  {journal} {JHEP}\ }\textbf {\bibinfo {volume} {03}},\ \bibinfo
  {pages} {102} (\bibinfo {year} {2018})},\ \Eprint
  {http://arxiv.org/abs/1711.11360} {arXiv:1711.11360 [hep-ph]} \BibitemShut
  {NoStop}%
\bibitem [{\citenamefont {Golec-Biernat}\ and\ \citenamefont
  {Wusthoff}(1998)}]{Golec-Biernat:1998zce}%
  \BibitemOpen
  \bibfield  {author} {\bibinfo {author} {\bibfnamefont {K.~J.}\ \bibnamefont
  {Golec-Biernat}}\ and\ \bibinfo {author} {\bibfnamefont {M.}~\bibnamefont
  {Wusthoff}},\ }\href {\doibase 10.1103/PhysRevD.59.014017} {\bibfield
  {journal} {\bibinfo  {journal} {Phys. Rev. D}\ }\textbf {\bibinfo {volume}
  {59}},\ \bibinfo {pages} {014017} (\bibinfo {year} {1998})},\ \Eprint
  {http://arxiv.org/abs/hep-ph/9807513} {arXiv:hep-ph/9807513} \BibitemShut
  {NoStop}%
\bibitem [{\citenamefont {Golec-Biernat}\ and\ \citenamefont
  {Wusthoff}(1999)}]{Golec-Biernat:1999qor}%
  \BibitemOpen
  \bibfield  {author} {\bibinfo {author} {\bibfnamefont {K.~J.}\ \bibnamefont
  {Golec-Biernat}}\ and\ \bibinfo {author} {\bibfnamefont {M.}~\bibnamefont
  {Wusthoff}},\ }\href {\doibase 10.1103/PhysRevD.60.114023} {\bibfield
  {journal} {\bibinfo  {journal} {Phys. Rev. D}\ }\textbf {\bibinfo {volume}
  {60}},\ \bibinfo {pages} {114023} (\bibinfo {year} {1999})},\ \Eprint
  {http://arxiv.org/abs/hep-ph/9903358} {arXiv:hep-ph/9903358} \BibitemShut
  {NoStop}%
\bibitem [{\citenamefont {Timneanu}\ \emph {et~al.}(2002)\citenamefont
  {Timneanu}, \citenamefont {Kwiecinski},\ and\ \citenamefont
  {Motyka}}]{Timneanu:2001bk}%
  \BibitemOpen
  \bibfield  {author} {\bibinfo {author} {\bibfnamefont {N.}~\bibnamefont
  {Timneanu}}, \bibinfo {author} {\bibfnamefont {J.}~\bibnamefont
  {Kwiecinski}}, \ and\ \bibinfo {author} {\bibfnamefont {L.}~\bibnamefont
  {Motyka}},\ }\href {\doibase 10.1007/s100520200893} {\bibfield  {journal}
  {\bibinfo  {journal} {Eur. Phys. J. C}\ }\textbf {\bibinfo {volume} {23}},\
  \bibinfo {pages} {513} (\bibinfo {year} {2002})},\ \Eprint
  {http://arxiv.org/abs/hep-ph/0110409} {arXiv:hep-ph/0110409} \BibitemShut
  {NoStop}%
\bibitem [{\citenamefont {Forshaw}\ \emph {et~al.}(2004)\citenamefont
  {Forshaw}, \citenamefont {Sandapen},\ and\ \citenamefont
  {Shaw}}]{Forshaw:2003ki}%
  \BibitemOpen
  \bibfield  {author} {\bibinfo {author} {\bibfnamefont {J.~R.}\ \bibnamefont
  {Forshaw}}, \bibinfo {author} {\bibfnamefont {R.}~\bibnamefont {Sandapen}}, \
  and\ \bibinfo {author} {\bibfnamefont {G.}~\bibnamefont {Shaw}},\ }\href
  {\doibase 10.1103/PhysRevD.69.094013} {\bibfield  {journal} {\bibinfo
  {journal} {Phys. Rev. D}\ }\textbf {\bibinfo {volume} {69}},\ \bibinfo
  {pages} {094013} (\bibinfo {year} {2004})},\ \Eprint
  {http://arxiv.org/abs/hep-ph/0312172} {arXiv:hep-ph/0312172} \BibitemShut
  {NoStop}%
\bibitem [{\citenamefont {Flensburg}\ \emph {et~al.}(2009)\citenamefont
  {Flensburg}, \citenamefont {Gustafson},\ and\ \citenamefont
  {Lonnblad}}]{Flensburg:2008ag}%
  \BibitemOpen
  \bibfield  {author} {\bibinfo {author} {\bibfnamefont {C.}~\bibnamefont
  {Flensburg}}, \bibinfo {author} {\bibfnamefont {G.}~\bibnamefont
  {Gustafson}}, \ and\ \bibinfo {author} {\bibfnamefont {L.}~\bibnamefont
  {Lonnblad}},\ }\href {\doibase 10.1140/epjc/s10052-009-0868-7} {\bibfield
  {journal} {\bibinfo  {journal} {Eur. Phys. J. C}\ }\textbf {\bibinfo {volume}
  {60}},\ \bibinfo {pages} {233} (\bibinfo {year} {2009})},\ \Eprint
  {http://arxiv.org/abs/0807.0325} {arXiv:0807.0325 [hep-ph]} \BibitemShut
  {NoStop}%
\bibitem [{\citenamefont {Henkels}\ \emph {et~al.}(2020)\citenamefont
  {Henkels}, \citenamefont {de~Oliveira}, \citenamefont {Pasechnik},\ and\
  \citenamefont {Trebien}}]{Henkels:2020kju}%
  \BibitemOpen
  \bibfield  {author} {\bibinfo {author} {\bibfnamefont {C.}~\bibnamefont
  {Henkels}}, \bibinfo {author} {\bibfnamefont {E.~G.}\ \bibnamefont
  {de~Oliveira}}, \bibinfo {author} {\bibfnamefont {R.}~\bibnamefont
  {Pasechnik}}, \ and\ \bibinfo {author} {\bibfnamefont {H.}~\bibnamefont
  {Trebien}},\ }\href {\doibase 10.1103/PhysRevD.102.014024} {\bibfield
  {journal} {\bibinfo  {journal} {Phys. Rev. D}\ }\textbf {\bibinfo {volume}
  {102}},\ \bibinfo {pages} {014024} (\bibinfo {year} {2020})},\ \Eprint
  {http://arxiv.org/abs/2004.00607} {arXiv:2004.00607 [hep-ph]} \BibitemShut
  {NoStop}%
\bibitem [{\citenamefont {Goda}\ \emph {et~al.}(2023)\citenamefont {Goda},
  \citenamefont {Kutak},\ and\ \citenamefont {Sapeta}}]{Goda:2023jie}%
  \BibitemOpen
  \bibfield  {author} {\bibinfo {author} {\bibfnamefont {T.}~\bibnamefont
  {Goda}}, \bibinfo {author} {\bibfnamefont {K.}~\bibnamefont {Kutak}}, \ and\
  \bibinfo {author} {\bibfnamefont {S.}~\bibnamefont {Sapeta}},\ }\href
  {\doibase 10.1140/epjc/s10052-023-12064-y} {\bibfield  {journal} {\bibinfo
  {journal} {Eur. Phys. J. C}\ }\textbf {\bibinfo {volume} {83}},\ \bibinfo
  {pages} {957} (\bibinfo {year} {2023})},\ \Eprint
  {http://arxiv.org/abs/2305.14025} {arXiv:2305.14025 [hep-ph]} \BibitemShut
  {NoStop}%
\bibitem [{\citenamefont {Albacete}\ \emph {et~al.}(2011)\citenamefont
  {Albacete}, \citenamefont {Armesto}, \citenamefont {Milhano}, \citenamefont
  {Quiroga-Arias},\ and\ \citenamefont {Salgado}}]{Albacete:2010sy}%
  \BibitemOpen
  \bibfield  {author} {\bibinfo {author} {\bibfnamefont {J.~L.}\ \bibnamefont
  {Albacete}}, \bibinfo {author} {\bibfnamefont {N.}~\bibnamefont {Armesto}},
  \bibinfo {author} {\bibfnamefont {J.~G.}\ \bibnamefont {Milhano}}, \bibinfo
  {author} {\bibfnamefont {P.}~\bibnamefont {Quiroga-Arias}}, \ and\ \bibinfo
  {author} {\bibfnamefont {C.~A.}\ \bibnamefont {Salgado}},\ }\href {\doibase
  10.1140/epjc/s10052-011-1705-3} {\bibfield  {journal} {\bibinfo  {journal}
  {Eur. Phys. J. C}\ }\textbf {\bibinfo {volume} {71}},\ \bibinfo {pages}
  {1705} (\bibinfo {year} {2011})},\ \Eprint {http://arxiv.org/abs/1012.4408}
  {arXiv:1012.4408 [hep-ph]} \BibitemShut {NoStop}%
\bibitem [{\citenamefont {Goncalves}\ \emph {et~al.}(2017)\citenamefont
  {Goncalves}, \citenamefont {Kopeliovich}, \citenamefont {Nemchik},
  \citenamefont {Pasechnik},\ and\ \citenamefont
  {Potashnikova}}]{Goncalves:2017chx}%
  \BibitemOpen
  \bibfield  {author} {\bibinfo {author} {\bibfnamefont {V.~P.}\ \bibnamefont
  {Goncalves}}, \bibinfo {author} {\bibfnamefont {B.}~\bibnamefont
  {Kopeliovich}}, \bibinfo {author} {\bibfnamefont {J.}~\bibnamefont
  {Nemchik}}, \bibinfo {author} {\bibfnamefont {R.}~\bibnamefont {Pasechnik}},
  \ and\ \bibinfo {author} {\bibfnamefont {I.}~\bibnamefont {Potashnikova}},\
  }\href {\doibase 10.1103/PhysRevD.96.014010} {\bibfield  {journal} {\bibinfo
  {journal} {Phys. Rev. D}\ }\textbf {\bibinfo {volume} {96}},\ \bibinfo
  {pages} {014010} (\bibinfo {year} {2017})},\ \Eprint
  {http://arxiv.org/abs/1704.04699} {arXiv:1704.04699 [hep-ph]} \BibitemShut
  {NoStop}%
\bibitem [{\citenamefont {Luszczak}\ and\ \citenamefont
  {Kowalski}(2014)}]{Luszczak:2013rxa}%
  \BibitemOpen
  \bibfield  {author} {\bibinfo {author} {\bibfnamefont {A.}~\bibnamefont
  {Luszczak}}\ and\ \bibinfo {author} {\bibfnamefont {H.}~\bibnamefont
  {Kowalski}},\ }\href {\doibase 10.1103/PhysRevD.89.074051} {\bibfield
  {journal} {\bibinfo  {journal} {Phys. Rev. D}\ }\textbf {\bibinfo {volume}
  {89}},\ \bibinfo {pages} {074051} (\bibinfo {year} {2014})},\ \Eprint
  {http://arxiv.org/abs/1312.4060} {arXiv:1312.4060 [hep-ph]} \BibitemShut
  {NoStop}%
\bibitem [{\citenamefont {Bartels}\ \emph {et~al.}(2003)\citenamefont
  {Bartels}, \citenamefont {Gotsman}, \citenamefont {Levin}, \citenamefont
  {Lublinsky},\ and\ \citenamefont {Maor}}]{Bartels:2002uf}%
  \BibitemOpen
  \bibfield  {author} {\bibinfo {author} {\bibfnamefont {J.}~\bibnamefont
  {Bartels}}, \bibinfo {author} {\bibfnamefont {E.}~\bibnamefont {Gotsman}},
  \bibinfo {author} {\bibfnamefont {E.}~\bibnamefont {Levin}}, \bibinfo
  {author} {\bibfnamefont {M.}~\bibnamefont {Lublinsky}}, \ and\ \bibinfo
  {author} {\bibfnamefont {U.}~\bibnamefont {Maor}},\ }\href {\doibase
  10.1016/S0370-2693(03)00128-X} {\bibfield  {journal} {\bibinfo  {journal}
  {Phys. Lett. B}\ }\textbf {\bibinfo {volume} {556}},\ \bibinfo {pages} {114}
  (\bibinfo {year} {2003})},\ \Eprint {http://arxiv.org/abs/hep-ph/0212284}
  {arXiv:hep-ph/0212284} \BibitemShut {NoStop}%
\bibitem [{\citenamefont {Brodsky}\ and\ \citenamefont
  {de~Teramond}(2009)}]{Brodsky:2008pg}%
  \BibitemOpen
  \bibfield  {author} {\bibinfo {author} {\bibfnamefont {S.~J.}\ \bibnamefont
  {Brodsky}}\ and\ \bibinfo {author} {\bibfnamefont {G.~F.}\ \bibnamefont
  {de~Teramond}},\ }\href {\doibase 10.1142/9789814293242_0008} {\bibfield
  {journal} {\bibinfo  {journal} {Subnucl. Ser.}\ }\textbf {\bibinfo {volume}
  {45}},\ \bibinfo {pages} {139} (\bibinfo {year} {2009})},\ \Eprint
  {http://arxiv.org/abs/0802.0514} {arXiv:0802.0514 [hep-ph]} \BibitemShut
  {NoStop}%
\bibitem [{\citenamefont {Brodsky}\ \emph {et~al.}(2015)\citenamefont
  {Brodsky}, \citenamefont {de~Teramond}, \citenamefont {Dosch},\ and\
  \citenamefont {Erlich}}]{Brodsky:2014yha}%
  \BibitemOpen
  \bibfield  {author} {\bibinfo {author} {\bibfnamefont {S.~J.}\ \bibnamefont
  {Brodsky}}, \bibinfo {author} {\bibfnamefont {G.~F.}\ \bibnamefont
  {de~Teramond}}, \bibinfo {author} {\bibfnamefont {H.~G.}\ \bibnamefont
  {Dosch}}, \ and\ \bibinfo {author} {\bibfnamefont {J.}~\bibnamefont
  {Erlich}},\ }\href {\doibase 10.1016/j.physrep.2015.05.001} {\bibfield
  {journal} {\bibinfo  {journal} {Phys. Rept.}\ }\textbf {\bibinfo {volume}
  {584}},\ \bibinfo {pages} {1} (\bibinfo {year} {2015})},\ \Eprint
  {http://arxiv.org/abs/1407.8131} {arXiv:1407.8131 [hep-ph]} \BibitemShut
  {NoStop}%
\bibitem [{\citenamefont {Forshaw}\ and\ \citenamefont
  {Sandapen}(2012)}]{Forshaw:2012im}%
  \BibitemOpen
  \bibfield  {author} {\bibinfo {author} {\bibfnamefont {J.~R.}\ \bibnamefont
  {Forshaw}}\ and\ \bibinfo {author} {\bibfnamefont {R.}~\bibnamefont
  {Sandapen}},\ }\href {\doibase 10.1103/PhysRevLett.109.081601} {\bibfield
  {journal} {\bibinfo  {journal} {Phys. Rev. Lett.}\ }\textbf {\bibinfo
  {volume} {109}},\ \bibinfo {pages} {081601} (\bibinfo {year} {2012})},\
  \Eprint {http://arxiv.org/abs/1203.6088} {arXiv:1203.6088 [hep-ph]}
  \BibitemShut {NoStop}%
\bibitem [{\citenamefont {Kowalski}\ and\ \citenamefont
  {Teaney}(2003)}]{Kowalski:2003hm}%
  \BibitemOpen
  \bibfield  {author} {\bibinfo {author} {\bibfnamefont {H.}~\bibnamefont
  {Kowalski}}\ and\ \bibinfo {author} {\bibfnamefont {D.}~\bibnamefont
  {Teaney}},\ }\href {\doibase 10.1103/PhysRevD.68.114005} {\bibfield
  {journal} {\bibinfo  {journal} {Phys. Rev. D}\ }\textbf {\bibinfo {volume}
  {68}},\ \bibinfo {pages} {114005} (\bibinfo {year} {2003})},\ \Eprint
  {http://arxiv.org/abs/hep-ph/0304189} {arXiv:hep-ph/0304189} \BibitemShut
  {NoStop}%
\bibitem [{\citenamefont {Frankfurt}\ \emph
  {et~al.}(1998{\natexlab{a}})\citenamefont {Frankfurt}, \citenamefont
  {Koepf},\ and\ \citenamefont {Strikman}}]{Frankfurt:1997fj}%
  \BibitemOpen
  \bibfield  {author} {\bibinfo {author} {\bibfnamefont {L.}~\bibnamefont
  {Frankfurt}}, \bibinfo {author} {\bibfnamefont {W.}~\bibnamefont {Koepf}}, \
  and\ \bibinfo {author} {\bibfnamefont {M.}~\bibnamefont {Strikman}},\ }\href
  {\doibase 10.1103/PhysRevD.57.512} {\bibfield  {journal} {\bibinfo  {journal}
  {Phys. Rev. D}\ }\textbf {\bibinfo {volume} {57}},\ \bibinfo {pages} {512}
  (\bibinfo {year} {1998}{\natexlab{a}})},\ \Eprint
  {http://arxiv.org/abs/hep-ph/9702216} {arXiv:hep-ph/9702216} \BibitemShut
  {NoStop}%
\bibitem [{\citenamefont {Dirac}(1949)}]{RevModPhys.21.392}%
  \BibitemOpen
  \bibfield  {author} {\bibinfo {author} {\bibfnamefont {P.~A.~M.}\
  \bibnamefont {Dirac}},\ }\href {\doibase 10.1103/RevModPhys.21.392}
  {\bibfield  {journal} {\bibinfo  {journal} {Rev. Mod. Phys.}\ }\textbf
  {\bibinfo {volume} {21}},\ \bibinfo {pages} {392} (\bibinfo {year}
  {1949})}\BibitemShut {NoStop}%
\bibitem [{\citenamefont {Patrignani}\ \emph {et~al.}(2016)\citenamefont
  {Patrignani} \emph {et~al.}}]{ParticleDataGroup:2016lqr}%
  \BibitemOpen
  \bibfield  {author} {\bibinfo {author} {\bibfnamefont {C.}~\bibnamefont
  {Patrignani}} \emph {et~al.} (\bibinfo {collaboration} {Particle Data
  Group}),\ }\href {\doibase 10.1088/1674-1137/40/10/100001} {\bibfield
  {journal} {\bibinfo  {journal} {Chin. Phys. C}\ }\textbf {\bibinfo {volume}
  {40}},\ \bibinfo {pages} {100001} (\bibinfo {year} {2016})}\BibitemShut
  {NoStop}%
\bibitem [{\citenamefont {Frankfurt}\ \emph
  {et~al.}(1998{\natexlab{b}})\citenamefont {Frankfurt}, \citenamefont
  {Guzey},\ and\ \citenamefont {Strikman}}]{Frankfurt:1997zk}%
  \BibitemOpen
  \bibfield  {author} {\bibinfo {author} {\bibfnamefont {L.}~\bibnamefont
  {Frankfurt}}, \bibinfo {author} {\bibfnamefont {V.}~\bibnamefont {Guzey}}, \
  and\ \bibinfo {author} {\bibfnamefont {M.}~\bibnamefont {Strikman}},\ }\href
  {\doibase 10.1103/PhysRevD.58.094039} {\bibfield  {journal} {\bibinfo
  {journal} {Phys. Rev. D}\ }\textbf {\bibinfo {volume} {58}},\ \bibinfo
  {pages} {094039} (\bibinfo {year} {1998}{\natexlab{b}})},\ \Eprint
  {http://arxiv.org/abs/hep-ph/9712339} {arXiv:hep-ph/9712339} \BibitemShut
  {NoStop}%
\bibitem [{\citenamefont {Frankfurt}\ \emph {et~al.}(2016)\citenamefont
  {Frankfurt}, \citenamefont {Guzey}, \citenamefont {Strikman},\ and\
  \citenamefont {Zhalov}}]{Frankfurt:2015cwa}%
  \BibitemOpen
  \bibfield  {author} {\bibinfo {author} {\bibfnamefont {L.}~\bibnamefont
  {Frankfurt}}, \bibinfo {author} {\bibfnamefont {V.}~\bibnamefont {Guzey}},
  \bibinfo {author} {\bibfnamefont {M.}~\bibnamefont {Strikman}}, \ and\
  \bibinfo {author} {\bibfnamefont {M.}~\bibnamefont {Zhalov}},\ }\href
  {\doibase 10.1016/j.physletb.2015.11.012} {\bibfield  {journal} {\bibinfo
  {journal} {Phys. Lett. B}\ }\textbf {\bibinfo {volume} {752}},\ \bibinfo
  {pages} {51} (\bibinfo {year} {2016})},\ \Eprint
  {http://arxiv.org/abs/1506.07150} {arXiv:1506.07150 [hep-ph]} \BibitemShut
  {NoStop}%
\bibitem [{\citenamefont {Khoze}\ \emph {et~al.}(2019)\citenamefont {Khoze},
  \citenamefont {Martin},\ and\ \citenamefont {Ryskin}}]{Khoze:2019xke}%
  \BibitemOpen
  \bibfield  {author} {\bibinfo {author} {\bibfnamefont {V.~A.}\ \bibnamefont
  {Khoze}}, \bibinfo {author} {\bibfnamefont {A.~D.}\ \bibnamefont {Martin}}, \
  and\ \bibinfo {author} {\bibfnamefont {M.~G.}\ \bibnamefont {Ryskin}},\
  }\href {\doibase 10.1088/1361-6471/ab2009} {\bibfield  {journal} {\bibinfo
  {journal} {J. Phys. G}\ }\textbf {\bibinfo {volume} {46}},\ \bibinfo {pages}
  {085002} (\bibinfo {year} {2019})},\ \Eprint
  {http://arxiv.org/abs/1902.08136} {arXiv:1902.08136 [hep-ph]} \BibitemShut
  {NoStop}%
\bibitem [{\citenamefont {Ivanov}\ \emph {et~al.}(2002)\citenamefont {Ivanov},
  \citenamefont {Kopeliovich}, \citenamefont {Tarasov},\ and\ \citenamefont
  {Hufner}}]{Ivanov:2002kc}%
  \BibitemOpen
  \bibfield  {author} {\bibinfo {author} {\bibfnamefont {{\relax Yu}.~P.}\
  \bibnamefont {Ivanov}}, \bibinfo {author} {\bibfnamefont {B.~Z.}\
  \bibnamefont {Kopeliovich}}, \bibinfo {author} {\bibfnamefont {A.~V.}\
  \bibnamefont {Tarasov}}, \ and\ \bibinfo {author} {\bibfnamefont
  {J.}~\bibnamefont {Hufner}},\ }\href {\doibase 10.1103/PhysRevC.66.024903}
  {\bibfield  {journal} {\bibinfo  {journal} {Phys. Rev.}\ }\textbf {\bibinfo
  {volume} {C66}},\ \bibinfo {pages} {024903} (\bibinfo {year} {2002})},\
  \Eprint {http://arxiv.org/abs/hep-ph/0202216} {arXiv:hep-ph/0202216 [hep-ph]}
  \BibitemShut {NoStop}%
%%CITATION = HEP-PH/0202216;%%
\bibitem [{\citenamefont {Ivanov}\ \emph {et~al.}(2003)\citenamefont {Ivanov},
  \citenamefont {Kopeliovich}, \citenamefont {Tarasov},\ and\ \citenamefont
  {Hufner}}]{Ivanov:2002eq}%
  \BibitemOpen
  \bibfield  {author} {\bibinfo {author} {\bibfnamefont {Y.~P.}\ \bibnamefont
  {Ivanov}}, \bibinfo {author} {\bibfnamefont {B.~Z.}\ \bibnamefont
  {Kopeliovich}}, \bibinfo {author} {\bibfnamefont {A.~V.}\ \bibnamefont
  {Tarasov}}, \ and\ \bibinfo {author} {\bibfnamefont {J.}~\bibnamefont
  {Hufner}},\ }\bibfield  {booktitle} {\emph {\bibinfo {booktitle} {{Hadron
  physics: Effective theories of low energy QCD. Proceedings, 2nd International
  Workshop, Coimbra, Portugal, September 25-29, 2002}}},\ }\href {\doibase
  10.1063/1.1570580} {\bibfield  {journal} {\bibinfo  {journal} {AIP Conf.
  Proc.}\ }\textbf {\bibinfo {volume} {660}},\ \bibinfo {pages} {283} (\bibinfo
  {year} {2003})},\ \Eprint {http://arxiv.org/abs/hep-ph/0212322}
  {arXiv:hep-ph/0212322 [hep-ph]} \BibitemShut {NoStop}%
%%CITATION = HEP-PH/0212322;%%
\bibitem [{\citenamefont {Guzey}\ \emph {et~al.}(2016)\citenamefont {Guzey},
  \citenamefont {Kryshen},\ and\ \citenamefont {Zhalov}}]{Guzey:2016piu}%
  \BibitemOpen
  \bibfield  {author} {\bibinfo {author} {\bibfnamefont {V.}~\bibnamefont
  {Guzey}}, \bibinfo {author} {\bibfnamefont {E.}~\bibnamefont {Kryshen}}, \
  and\ \bibinfo {author} {\bibfnamefont {M.}~\bibnamefont {Zhalov}},\ }\href
  {\doibase 10.1103/PhysRevC.93.055206} {\bibfield  {journal} {\bibinfo
  {journal} {Phys. Rev. C}\ }\textbf {\bibinfo {volume} {93}},\ \bibinfo
  {pages} {055206} (\bibinfo {year} {2016})},\ \Eprint
  {http://arxiv.org/abs/1602.01456} {arXiv:1602.01456 [nucl-th]} \BibitemShut
  {NoStop}%
\bibitem [{\citenamefont {Cudell}\ \emph {et~al.}(2002)\citenamefont {Cudell},
  \citenamefont {Ezhela}, \citenamefont {Gauron}, \citenamefont {Kang},
  \citenamefont {Kuyanov}, \citenamefont {Lugovsky}, \citenamefont {Martynov},
  \citenamefont {Nicolescu}, \citenamefont {Razuvaev},\ and\ \citenamefont
  {Tkachenko}}]{COMPETE:2002jcr}%
  \BibitemOpen
  \bibfield  {author} {\bibinfo {author} {\bibfnamefont {J.~R.}\ \bibnamefont
  {Cudell}}, \bibinfo {author} {\bibfnamefont {V.~V.}\ \bibnamefont {Ezhela}},
  \bibinfo {author} {\bibfnamefont {P.}~\bibnamefont {Gauron}}, \bibinfo
  {author} {\bibfnamefont {K.}~\bibnamefont {Kang}}, \bibinfo {author}
  {\bibfnamefont {Y.~V.}\ \bibnamefont {Kuyanov}}, \bibinfo {author}
  {\bibfnamefont {S.~B.}\ \bibnamefont {Lugovsky}}, \bibinfo {author}
  {\bibfnamefont {E.}~\bibnamefont {Martynov}}, \bibinfo {author}
  {\bibfnamefont {B.}~\bibnamefont {Nicolescu}}, \bibinfo {author}
  {\bibfnamefont {E.~A.}\ \bibnamefont {Razuvaev}}, \ and\ \bibinfo {author}
  {\bibfnamefont {N.~P.}\ \bibnamefont {Tkachenko}} (\bibinfo {collaboration}
  {COMPETE}),\ }\href {\doibase 10.1103/PhysRevLett.89.201801} {\bibfield
  {journal} {\bibinfo  {journal} {Phys. Rev. Lett.}\ }\textbf {\bibinfo
  {volume} {89}},\ \bibinfo {pages} {201801} (\bibinfo {year} {2002})},\
  \Eprint {http://arxiv.org/abs/hep-ph/0206172} {arXiv:hep-ph/0206172}
  \BibitemShut {NoStop}%
\bibitem [{\citenamefont {von Weizsacker}(1934)}]{vonWeizsacker:1934nji}%
  \BibitemOpen
  \bibfield  {author} {\bibinfo {author} {\bibfnamefont {C.~F.}\ \bibnamefont
  {von Weizsacker}},\ }\href {\doibase 10.1007/BF01333110} {\bibfield
  {journal} {\bibinfo  {journal} {Z. Phys.}\ }\textbf {\bibinfo {volume}
  {88}},\ \bibinfo {pages} {612} (\bibinfo {year} {1934})}\BibitemShut
  {NoStop}%
%%CITATION = ZEPYA,88,612;%%
\bibitem [{\citenamefont {Williams}(1934)}]{Williams:1934ad}%
  \BibitemOpen
  \bibfield  {author} {\bibinfo {author} {\bibfnamefont {E.~J.}\ \bibnamefont
  {Williams}},\ }\href {\doibase 10.1103/PhysRev.45.729} {\bibfield  {journal}
  {\bibinfo  {journal} {Phys. Rev.}\ }\textbf {\bibinfo {volume} {45}},\
  \bibinfo {pages} {729} (\bibinfo {year} {1934})}\BibitemShut {NoStop}%
%%CITATION = PHRVA,45,729;%%
\bibitem [{\citenamefont {Woods}\ and\ \citenamefont
  {Saxon}(1954)}]{Woods:1954zz}%
  \BibitemOpen
  \bibfield  {author} {\bibinfo {author} {\bibfnamefont {R.~D.}\ \bibnamefont
  {Woods}}\ and\ \bibinfo {author} {\bibfnamefont {D.~S.}\ \bibnamefont
  {Saxon}},\ }\href {\doibase 10.1103/PhysRev.95.577} {\bibfield  {journal}
  {\bibinfo  {journal} {Phys. Rev.}\ }\textbf {\bibinfo {volume} {95}},\
  \bibinfo {pages} {577} (\bibinfo {year} {1954})}\BibitemShut {NoStop}%
%%CITATION = PHRVA,95,577;%%
\bibitem [{\citenamefont {Euteneuer}\ \emph {et~al.}(1978)\citenamefont
  {Euteneuer}, \citenamefont {Friedrich},\ and\ \citenamefont
  {Vogler}}]{Euteneuer:1978qw}%
  \BibitemOpen
  \bibfield  {author} {\bibinfo {author} {\bibfnamefont {H.}~\bibnamefont
  {Euteneuer}}, \bibinfo {author} {\bibfnamefont {J.}~\bibnamefont
  {Friedrich}}, \ and\ \bibinfo {author} {\bibfnamefont {N.}~\bibnamefont
  {Vogler}},\ }\href {\doibase 10.1016/0375-9474(78)90143-4} {\bibfield
  {journal} {\bibinfo  {journal} {Nucl. Phys.}\ }\textbf {\bibinfo {volume}
  {A298}},\ \bibinfo {pages} {452} (\bibinfo {year} {1978})}\BibitemShut
  {NoStop}%
%%CITATION = NUPHA,A298,452;%%
\bibitem [{\citenamefont {Kogut}\ and\ \citenamefont
  {Soper}(1970)}]{Kogut:1969xa}%
  \BibitemOpen
  \bibfield  {author} {\bibinfo {author} {\bibfnamefont {J.~B.}\ \bibnamefont
  {Kogut}}\ and\ \bibinfo {author} {\bibfnamefont {D.~E.}\ \bibnamefont
  {Soper}},\ }\href {\doibase 10.1103/PhysRevD.1.2901} {\bibfield  {journal}
  {\bibinfo  {journal} {Phys. Rev. D}\ }\textbf {\bibinfo {volume} {1}},\
  \bibinfo {pages} {2901} (\bibinfo {year} {1970})}\BibitemShut {NoStop}%
\bibitem [{\citenamefont {Bjorken}\ \emph {et~al.}(1971)\citenamefont
  {Bjorken}, \citenamefont {Kogut},\ and\ \citenamefont
  {Soper}}]{Bjorken:1970ah}%
  \BibitemOpen
  \bibfield  {author} {\bibinfo {author} {\bibfnamefont {J.}~\bibnamefont
  {Bjorken}}, \bibinfo {author} {\bibfnamefont {J.~B.}\ \bibnamefont {Kogut}},
  \ and\ \bibinfo {author} {\bibfnamefont {D.~E.}\ \bibnamefont {Soper}},\
  }\href {\doibase 10.1103/PhysRevD.3.1382} {\bibfield  {journal} {\bibinfo
  {journal} {Phys. Rev. D}\ }\textbf {\bibinfo {volume} {3}},\ \bibinfo {pages}
  {1382} (\bibinfo {year} {1971})}\BibitemShut {NoStop}%
\bibitem [{\citenamefont {Dosch}\ \emph {et~al.}(1997)\citenamefont {Dosch},
  \citenamefont {Gousset}, \citenamefont {Kulzinger},\ and\ \citenamefont
  {Pirner}}]{Dosch:1996ss}%
  \BibitemOpen
  \bibfield  {author} {\bibinfo {author} {\bibfnamefont {H.~G.}\ \bibnamefont
  {Dosch}}, \bibinfo {author} {\bibfnamefont {T.}~\bibnamefont {Gousset}},
  \bibinfo {author} {\bibfnamefont {G.}~\bibnamefont {Kulzinger}}, \ and\
  \bibinfo {author} {\bibfnamefont {H.~J.}\ \bibnamefont {Pirner}},\ }\href
  {\doibase 10.1103/PhysRevD.55.2602} {\bibfield  {journal} {\bibinfo
  {journal} {Phys. Rev. D}\ }\textbf {\bibinfo {volume} {55}},\ \bibinfo
  {pages} {2602} (\bibinfo {year} {1997})},\ \Eprint
  {http://arxiv.org/abs/hep-ph/9608203} {arXiv:hep-ph/9608203} \BibitemShut
  {NoStop}%
\bibitem [{\citenamefont {Lepage}\ and\ \citenamefont
  {Brodsky}(1980)}]{Lepage:1980fj}%
  \BibitemOpen
  \bibfield  {author} {\bibinfo {author} {\bibfnamefont {G.~P.}\ \bibnamefont
  {Lepage}}\ and\ \bibinfo {author} {\bibfnamefont {S.~J.}\ \bibnamefont
  {Brodsky}},\ }\href {\doibase 10.1103/PhysRevD.22.2157} {\bibfield  {journal}
  {\bibinfo  {journal} {Phys. Rev. D}\ }\textbf {\bibinfo {volume} {22}},\
  \bibinfo {pages} {2157} (\bibinfo {year} {1980})}\BibitemShut {NoStop}%
\bibitem [{\citenamefont {Kowalski}\ \emph {et~al.}(2006)\citenamefont
  {Kowalski}, \citenamefont {Motyka},\ and\ \citenamefont
  {Watt}}]{Kowalski:2006hc}%
  \BibitemOpen
  \bibfield  {author} {\bibinfo {author} {\bibfnamefont {H.}~\bibnamefont
  {Kowalski}}, \bibinfo {author} {\bibfnamefont {L.}~\bibnamefont {Motyka}}, \
  and\ \bibinfo {author} {\bibfnamefont {G.}~\bibnamefont {Watt}},\ }\href
  {\doibase 10.1103/PhysRevD.74.074016} {\bibfield  {journal} {\bibinfo
  {journal} {Phys. Rev.}\ }\textbf {\bibinfo {volume} {D74}},\ \bibinfo {pages}
  {074016} (\bibinfo {year} {2006})},\ \Eprint
  {http://arxiv.org/abs/hep-ph/0606272} {arXiv:hep-ph/0606272 [hep-ph]}
  \BibitemShut {NoStop}%
%%CITATION = HEP-PH/0606272;%%
\bibitem [{\citenamefont {Gribov}(1969)}]{Gribov:1968jf}%
  \BibitemOpen
  \bibfield  {author} {\bibinfo {author} {\bibfnamefont {V.~N.}\ \bibnamefont
  {Gribov}},\ }\href@noop {} {\bibfield  {journal} {\bibinfo  {journal} {Sov.
  Phys. JETP}\ }\textbf {\bibinfo {volume} {29}},\ \bibinfo {pages} {483}
  (\bibinfo {year} {1969})}\BibitemShut {NoStop}%
\bibitem [{\citenamefont {Shuvaev}\ \emph {et~al.}(1999)\citenamefont
  {Shuvaev}, \citenamefont {Golec-Biernat}, \citenamefont {Martin},\ and\
  \citenamefont {Ryskin}}]{Shuvaev:1999ce}%
  \BibitemOpen
  \bibfield  {author} {\bibinfo {author} {\bibfnamefont {A.~G.}\ \bibnamefont
  {Shuvaev}}, \bibinfo {author} {\bibfnamefont {K.~J.}\ \bibnamefont
  {Golec-Biernat}}, \bibinfo {author} {\bibfnamefont {A.~D.}\ \bibnamefont
  {Martin}}, \ and\ \bibinfo {author} {\bibfnamefont {M.~G.}\ \bibnamefont
  {Ryskin}},\ }\href {\doibase 10.1103/PhysRevD.60.014015} {\bibfield
  {journal} {\bibinfo  {journal} {Phys. Rev. D}\ }\textbf {\bibinfo {volume}
  {60}},\ \bibinfo {pages} {014015} (\bibinfo {year} {1999})},\ \Eprint
  {http://arxiv.org/abs/hep-ph/9902410} {arXiv:hep-ph/9902410} \BibitemShut
  {NoStop}%
\bibitem [{\citenamefont {Kopeliovich}\ \emph {et~al.}(2023)\citenamefont
  {Kopeliovich}, \citenamefont {Krelina}, \citenamefont {Nemchik},\ and\
  \citenamefont {Potashnikova}}]{Kopeliovich:2020has}%
  \BibitemOpen
  \bibfield  {author} {\bibinfo {author} {\bibfnamefont {B.~Z.}\ \bibnamefont
  {Kopeliovich}}, \bibinfo {author} {\bibfnamefont {M.}~\bibnamefont
  {Krelina}}, \bibinfo {author} {\bibfnamefont {J.}~\bibnamefont {Nemchik}}, \
  and\ \bibinfo {author} {\bibfnamefont {I.~K.}\ \bibnamefont {Potashnikova}},\
  }\href {\doibase 10.1103/PhysRevD.107.054005} {\bibfield  {journal} {\bibinfo
   {journal} {Phys. Rev. D}\ }\textbf {\bibinfo {volume} {107}},\ \bibinfo
  {pages} {054005} (\bibinfo {year} {2023})},\ \Eprint
  {http://arxiv.org/abs/2008.05116} {arXiv:2008.05116 [hep-ph]} \BibitemShut
  {NoStop}%
\bibitem [{\citenamefont {Kopeliovich}\ \emph {et~al.}(2022)\citenamefont
  {Kopeliovich}, \citenamefont {Krelina}, \citenamefont {Nemchik},\ and\
  \citenamefont {Potashnikova}}]{Kopeliovich:2022jwe}%
  \BibitemOpen
  \bibfield  {author} {\bibinfo {author} {\bibfnamefont {B.~Z.}\ \bibnamefont
  {Kopeliovich}}, \bibinfo {author} {\bibfnamefont {M.}~\bibnamefont
  {Krelina}}, \bibinfo {author} {\bibfnamefont {J.}~\bibnamefont {Nemchik}}, \
  and\ \bibinfo {author} {\bibfnamefont {I.~K.}\ \bibnamefont {Potashnikova}},\
  }\href {\doibase 10.1103/PhysRevD.105.054023} {\bibfield  {journal} {\bibinfo
   {journal} {Phys. Rev. D}\ }\textbf {\bibinfo {volume} {105}},\ \bibinfo
  {pages} {054023} (\bibinfo {year} {2022})},\ \Eprint
  {http://arxiv.org/abs/2201.13021} {arXiv:2201.13021 [hep-ph]} \BibitemShut
  {NoStop}%
\bibitem [{\citenamefont {Adams}\ \emph {et~al.}(1995)\citenamefont {Adams}
  \emph {et~al.}}]{E665:1995xur}%
  \BibitemOpen
  \bibfield  {author} {\bibinfo {author} {\bibfnamefont {M.~R.}\ \bibnamefont
  {Adams}} \emph {et~al.} (\bibinfo {collaboration} {E665}),\ }\href {\doibase
  10.1007/BF01624583} {\bibfield  {journal} {\bibinfo  {journal} {Z. Phys. C}\
  }\textbf {\bibinfo {volume} {67}},\ \bibinfo {pages} {403} (\bibinfo {year}
  {1995})},\ \Eprint {http://arxiv.org/abs/hep-ex/9505006}
  {arXiv:hep-ex/9505006} \BibitemShut {NoStop}%
\bibitem [{\citenamefont {Armesto}(2002)}]{Armesto:2002ny}%
  \BibitemOpen
  \bibfield  {author} {\bibinfo {author} {\bibfnamefont {N.}~\bibnamefont
  {Armesto}},\ }\href {\doibase 10.1007/s10052-002-1021-z} {\bibfield
  {journal} {\bibinfo  {journal} {Eur. Phys. J. C}\ }\textbf {\bibinfo {volume}
  {26}},\ \bibinfo {pages} {35} (\bibinfo {year} {2002})},\ \Eprint
  {http://arxiv.org/abs/hep-ph/0206017} {arXiv:hep-ph/0206017} \BibitemShut
  {NoStop}%
\bibitem [{\citenamefont {Krelina}\ and\ \citenamefont
  {Nemchik}(2020)}]{Krelina:2020ipn}%
  \BibitemOpen
  \bibfield  {author} {\bibinfo {author} {\bibfnamefont {M.}~\bibnamefont
  {Krelina}}\ and\ \bibinfo {author} {\bibfnamefont {J.}~\bibnamefont
  {Nemchik}},\ }\href {\doibase 10.1140/epjp/s13360-020-00498-2} {\bibfield
  {journal} {\bibinfo  {journal} {Eur. Phys. J. Plus}\ }\textbf {\bibinfo
  {volume} {135}},\ \bibinfo {pages} {444} (\bibinfo {year} {2020})},\ \Eprint
  {http://arxiv.org/abs/2003.04156} {arXiv:2003.04156 [hep-ph]} \BibitemShut
  {NoStop}%
\bibitem [{\citenamefont {Kopeliovich}\ \emph {et~al.}(1998)\citenamefont
  {Kopeliovich}, \citenamefont {Raufeisen},\ and\ \citenamefont
  {Tarasov}}]{Kopeliovich:1998gv}%
  \BibitemOpen
  \bibfield  {author} {\bibinfo {author} {\bibfnamefont {B.~Z.}\ \bibnamefont
  {Kopeliovich}}, \bibinfo {author} {\bibfnamefont {J.}~\bibnamefont
  {Raufeisen}}, \ and\ \bibinfo {author} {\bibfnamefont {A.~V.}\ \bibnamefont
  {Tarasov}},\ }\href {\doibase 10.1016/S0370-2693(98)01072-7} {\bibfield
  {journal} {\bibinfo  {journal} {Phys. Lett. B}\ }\textbf {\bibinfo {volume}
  {440}},\ \bibinfo {pages} {151} (\bibinfo {year} {1998})},\ \Eprint
  {http://arxiv.org/abs/hep-ph/9807211} {arXiv:hep-ph/9807211} \BibitemShut
  {NoStop}%
\bibitem [{\citenamefont {Sirunyan}\ \emph {et~al.}(2019)\citenamefont
  {Sirunyan} \emph {et~al.}}]{CMS:2019awk}%
  \BibitemOpen
  \bibfield  {author} {\bibinfo {author} {\bibfnamefont {A.~M.}\ \bibnamefont
  {Sirunyan}} \emph {et~al.} (\bibinfo {collaboration} {CMS}),\ }\href
  {\doibase 10.1140/epjc/s10052-019-7202-9} {\bibfield  {journal} {\bibinfo
  {journal} {Eur. Phys. J. C}\ }\textbf {\bibinfo {volume} {79}},\ \bibinfo
  {pages} {702} (\bibinfo {year} {2019})},\ \Eprint
  {http://arxiv.org/abs/1902.01339} {arXiv:1902.01339 [hep-ex]} \BibitemShut
  {NoStop}%
\bibitem [{\citenamefont {Acharya}\ \emph {et~al.}(2021)\citenamefont {Acharya}
  \emph {et~al.}}]{ALICE:2021jnv}%
  \BibitemOpen
  \bibfield  {author} {\bibinfo {author} {\bibfnamefont {S.}~\bibnamefont
  {Acharya}} \emph {et~al.} (\bibinfo {collaboration} {ALICE}),\ }\href
  {\doibase 10.1016/j.physletb.2021.136481} {\bibfield  {journal} {\bibinfo
  {journal} {Phys. Lett. B}\ }\textbf {\bibinfo {volume} {820}},\ \bibinfo
  {pages} {136481} (\bibinfo {year} {2021})},\ \Eprint
  {http://arxiv.org/abs/2101.02581} {arXiv:2101.02581 [nucl-ex]} \BibitemShut
  {NoStop}%
\end{thebibliography}%

\end{document}